\documentclass[final,12pt]{article}
\usepackage{amsfonts,color,morefloats,pslatex,a4wide}
\usepackage{amssymb,amsthm,amsmath,latexsym,pslatex,cite}
\usepackage{lineno,hyperref,mathtools}
\usepackage[ruled,linesnumbered]{algorithm2e}
\usepackage{pdflscape}
\usepackage{microtype}
\usepackage{setspace}

\newtheorem{theorem}{Theorem}[section]

\newtheorem{lemma}[theorem]{Lemma}

\theoremstyle{definition}
\newtheorem{definition}[theorem]{Definition}
\newtheorem{example}[theorem]{Example}

\newtheorem{problem}[theorem]{Problem}

\newtheorem{remark}[theorem]{Remark}

\usepackage{soul}

\newcommand{\F}{\mathbb{F}}

\newcommand{\M}{\mathcal{M}}

\newcommand{\C}{{\mathcal{C}}}

\newcommand{\diag}{{\mathrm{{diag}}}}
\newcommand{\adiag}{{\mathrm{{adiag}}}}

\newcommand{\Aut}{{\mathbf{Aut}}}

\newcommand{\MAut}{{\mathbf{MAut}}}
\newcommand{\SLAut}{\mathbf{SLAut}}

\makeatletter

\newcommand{\Rmnum}[1]{\expandafter\@slowromancap\romannumeral #1@}
\makeatother

\begin{document}
\begin{sloppypar}

\title{On {general} self-orthogonal matrix-product codes associated with Toeplitz matrices\thanks{Corresponding author: Shixin Zhu.}}

\author{Yang Li\thanks{School of Mathematics, Hefei University of Technology, Hefei, 230601, Anhui, China. Email:  yanglimath@163.com.},
\and Shixin Zhu\thanks{School of Mathematics, Hefei University of Technology, Hefei, 230601, Anhui, China. Email: zhushixinmath@hfut.edu.cn.}, 
\and Edgar Mart\'inez-Moro\thanks{Institute of Mathematics, University of Valladolid, Spain. Email: Edgar.Martinez@uva.es.} 
}

\maketitle

\begin{abstract}
In this paper, we present four constructions of {general} self-orthogonal matrix-product codes associated with Toeplitz matrices. 
The first one relies on the {dual} of a known {general} dual-containing matrix-product code; 
the second one is founded on {a specific family of} matrices, where we provide an efficient algorithm 
for generating them {on the basis of Toeplitz matrices} and {it has an interesting application in producing 
new non-singular by columns quasi-unitary matrices};
and the last two ones are based on the utilization of certain special Toeplitz matrices. 
Concrete examples and detailed comparisons are provided. 
As a byproduct, we also find an application of Toeplitz matrices, which is closely related to the constructions of quantum codes. 
\end{abstract}

\noindent{\bf Keywords:} Self-orthogonal code, Matrix-product code, Toeplitz matrix, General construction \\
\noindent{\bf MSC(2010)}: Primary 94B05 15B05 12E10

\section{Introduction}\label{sec1}

Let $\F_q$ denote the finite field with $q$ elements, where $q=p^h$ is a prime power and 
let $\F_q^{n}$ denote the $n$-dimensional vector space over $\F_q$. 
Then an $[n,k,d]_q$ linear code $\C$ is a $k$-dimensional subspace of $\F_q^n$ with minimum Hamming distance $d$ 
and we call $\C$ a {\em maximum distance separable (MDS)} code if $d=n-k+1$. 
For a given linear code $\C$ over $\F_q$, we will use $\C^{\perp}$ to denote its {\em dual code} 
with respect to a certain inner product, such as the Euclidean, Hermitian, Galois, symplectic or $\sigma$ inner product. 
A linear code $\C$ is said to be 
{\em self-orthogonal} if $\C\subseteq \C^{\perp}$;  
{\em dual-containing} if $\C^{\perp}\subseteq \C$, 
and {\em self-dual} if $\C=\C^{\perp}$.

Since the very beginning of coding theory, the exploration of self-orthogonal codes has garnered 
significant interest and emerged as a focal point. 
Extensive research on this topic has revealed robust connections between self-orthogonal codes and 
diverse mathematical domains, including combinatorial $t$-design theory \cite{t-design}, group theory \cite{lattice2}, 
lattice theory \cite{lattice1,lattice2,lattice3}, modular forms \cite{modular}, and quantum coding theory \cite{quantum1,quantum2}. 
Specifically speaking, finite groups such as the Mathieu groups were linked to certain self-orthogonal codes, 
and the extended binary self-orthogonal Golay code was associated with the Conway group. 
Additionally, self-orthogonal codes have been used to produce many $5$-designs \cite{5-design}. 
For more details on self-orthogonal codes, one can refer to the recent papers 
\cite{LL2022,ZKY2023,SLK2023,LZ2024,C2023,HLL2023,WL2020,KKL2021,BV2022,XD2022} and the references therein. 
All these distinguished works have also  stimulated the interest in the study of self-orthogonal codes under various inner products.

On the other hand, matrix-product codes were introduced by Blackmore and Norton as a generalization of many combinatorial constructions 
 in \cite{BN2001} and hence, they provide efficient ways to obtain new codes of larger 
lengths from known codes of short lengths. 
Note also that efficient decoding algorithms for matrix-product codes were developed in \cite{HHR2012,HLR2009,HR2013}. 
Nowadays, matrix-product codes have been widely studied and applied in various fields, 
such as locally recoverable codes \cite{LEL2023LRC}, symbol-pair codes \cite{LELP2023symbol}, 
linear complementary pairs of codes \cite{LL2022}, and digital nets \cite{NO2004}.  
These results have also motivated us to further investigate the topic of matrix-product codes.



Combining the above two aspects, we have as a broad line of interest to study self-orthogonal matrix-product codes. 
Moreover, we have in mind for this research the following three motivations. 
\begin{itemize}
    \item [\rm (1)] Since the Euclidean and Hermitian duals of matrix-product codes were determined in \cite{BN2001,ZG2015},  
    many constructions of Euclidean and Hermitian self-orthogonal matrix-product codes have been proposed 
    in \cite{C2024,MJ2016,Z2024,JM2017,FLL2014} and the references therein. 
    Very recently, Cao $et~al.$ characterized the $\sigma$ duals of matrix-product codes in \cite{CYW2023}, 
    where the $\sigma$ dual is a generalization of the Euclidean and Hermitian duals. 
    On this basis, the authors obtained several constructions of $\sigma$ dual-containing matrix-product codes, 
    but none of $\sigma$ self-orthogonal matrix-product codes in the same paper. 
    Note also that as a natural generalization of Euclidean and Hermitian self-orthogonal codes, 
    when considering the $\sigma$ self-orthogonality, 
    it seems that only constacyclic codes over rings were studied in \cite{ZKY2023,LL2022sigma}. 
    In other words, it is still open to give several constructions of $\sigma$ self-orthogonal matrix-product codes over finite fields.
    
    \item [\rm (2)] {The constructions of Hermitian (resp. Euclidean) self-orthogonal or dual-containing matrix-product codes 
    such that they have good minimum distance lower bounds are closely related to non-singular by columns (NSC) quasi-unitary (resp. NSC quasi-orthogonal) matrices, 
    as reflected in \cite{CWC2020,JM2017,CC2020QIP,CC2020FFA}. 
    To achieve the constructions of these types of matrix-product codes, the authors in \cite{CWC2020,CW2024} studied the constructions of NSC quasi-unitary matrices 
    from NSC matrices and reversely NSC (RNSC) matrices, respectively, which involve non-singular Vandermonde matrices.}
    Note that Toeplitz matrices have recently gained substantial prominence in coding theory 
    as a generalization of circulant matrices and negacirculant matrices \cite{SXS2023,LZM2023,LLZ2024,LSL2023,LSW2023},  
    and these matrices are conveniently storable and computationally efficient 
    \cite{H1996Toeplitz2,M2011Toeplitz3,P2007Toeplitz1,MM1988Toeplitz4}. 
    Moreover, by our Magma \cite{magma} experiments, NSC matrices can be derived from Toeplitz matrices.
    This leads us to study the matrix-product codes related to Toeplitz matrices.

    \item [\rm (3)]  {As a general framework, it seems natural to take specific settings 
    in the constructions of $\sigma$ self-orthogonal matrix-product codes to 
    yield Euclidean and Hermitian self-orthogonal matrix-product codes \cite{ZKY2023,LL2022sigma}, 
    which have many applications in coding theory as we mentioned before. 
    As also pointed out in \cite[Remark 4.11 and Example 4.12]{CYW2023}, in some cases one cannot even 
    obtain appropriate NSC matrices for constructing Euclidean and Hermitian self-orthogonal matrix-product codes. 
    Conversely, it is still possible to obtain $\sigma$ self-orthogonal matrix-product codes in these cases.
    Furthermore, Cao recently developed the application of $\sigma$ self-orthogonal codes 
    in constructing entanglement-assisted quantum codes \cite{C2024Arxiv}. 
    However, once again, there is still no explicit construction of $\sigma$ self-orthogonal matrix-product codes.
    }
\end{itemize}

Motivated by the above three observations, it is of particular interest to study $\sigma$ self-orthogonal matrix-product codes 
associated with Toeplitz matrices. 
Focusing on this topic, four general constructions of $\sigma$ self-orthogonal matrix-product codes are presented. 
The following are the main results of this paper.

\begin{itemize}
    \item [\rm (1)] The first general construction of $\sigma$ self-orthogonal matrix-product codes, 
    presented in Theorem \ref{th.con1},  
    relies on the $\sigma'$ dual of a known $\sigma'$ dual-containing matrix-product code. 
    From it, we further state the relationship between a linear code and 
    its dual code with respect to a certain inner product in Remark \ref{rem.th.con1.222}. 

    \item [\rm (2)] We introduce NSC quasi-$\widehat{\sigma}$ matrices in Definition \ref{def.quasi-sigma matrix} 
    and present a concrete construction for them in Theorem \ref{th.NSC leading principalminors}. 
    An algorithm for generating NSC quasi-$\widehat{\sigma}$ matrices is further provided in Algorithm \ref{alg.NSC quasi-sigma matrix} 
    by employing Toeplitz matrices.  
    {We show some examples in Table \ref{tab2} to illustrate that Algorithm \ref{alg.NSC quasi-sigma matrix} is efficient 
    and it has an interesting application in deriving new NSC quasi-unitary matrices}. 
    With the help of these NSC quasi-$\widehat{\sigma}$ matrices, we obtain the second general construction of $\sigma$ self-orthogonal 
    matrix-product codes in Theorem \ref{th.con2}.

    \item [\rm (3)] The last two general constructions of $\sigma$ self-orthogonal matrix-product codes are
    presented in Theorems \ref{th.con3} and \ref{th.con4} by using certain 
    special Toeplitz matrices. 
    As a byproduct, we also find an attractive connection between those Toeplitz matrices and 
    $\widetilde{\tau}$-optimal defining matrices in Remark \ref{rem.tau-OD}.  

\end{itemize}


This paper is organized as follows. 
Section \ref{sec2} reviews some useful notions and results. 
Section \ref{sec.construction} presents the four general constructions of $\sigma$ self-orthogonal matrix-product codes 
with many specific examples and provides detailed comparisons among them. 
Finally, Section \ref{sec4} concludes this paper.

\section{Preliminaries}\label{sec2}

In this paper, we will denote by $q=p^h$ a prime power and $e$ will be an integer with $0\leq e\leq h-1$. 
$\F_q^{n}$ will denote the $n$-dimensional vector space over $\F_q$ and 
$\M(\F_q,n)$ will denote the $\F_q$-linear space of $n\times n$ matrices over $\F_q$.  
For two vectors $\mathbf{u}=(u_1,u_2,\ldots,u_{n})$ and $\mathbf{v}=(v_1,v_2,\ldots,v_{n})$ in $\F_q^{n}$, 
we denote by $d_{H}(\mathbf{u},\mathbf{v})=\sharp \{i\mid u_i\neq v_i~{\rm for}~1\leq i\leq n\}$  their \emph{Hamming distance}. 
$\mathbf{0}$ and $O$ will denote the zero vector and the zero matrix, respectively, whose sizes are unspecified here and will depend on the context. 
In the following, we review and give some useful results on $\sigma$ inner products, 
$\sigma$ duals, Toeplitz matrices, non-singular by columns matrices, and matrix-product codes.

\subsection{$\sigma$ inner products and $\sigma$ duals}

Let $\sigma$ be a mapping from $\F_q^n$ to $\F_q^n$. The mapping $\sigma$ is said to be an {\em isometry} 
if $d_H(\sigma(\mathbf{u}),\sigma(\mathbf{v}))=d_H(\mathbf{u},\mathbf{v})$ for any $\mathbf{u}, 
\mathbf{v} \in \F_q^n$. 
Moreover, if $\sigma$ is linear, we call it a {\em linear isometry}.  
The group of all isometries on $\F_q^{n}$ will be denoted by $\Aut(\F_q^{n})$. 
Two linear codes $\C_1$ and $\C_2$ are called {\em isometric} if $\C_2=\sigma(\C_1)$ for some $\sigma\in \Aut(\F_q^{n})$.
Let $\MAut(\F_q^{n})$ be the {\em monomial group} consisted of the set of linear maps given by monomial matrices in $\M(\F_q,n)$.  
It can be easily checked that $\MAut(\F_q^{n})$ corresponds to the group of all linear isometries of $\F_q^n$.

As the authors proved in \cite{semilinear1,semilinear2}, when $n\geq 3$, 
isometries that map subspaces onto subspaces are exactly the {\em semilinear mappings} of the form 
\begin{align*}
    \begin{split}
        \sigma=(\tau,\pi):\  \F_q^{n} & \to \F_q^{n} \\
                             \mathbf{u} & \mapsto \tau(\pi(\mathbf{u})),
    \end{split}
\end{align*}
where $\tau$ is a linear isometry and $\pi$ is an automorphism of the finite field $\F_q$, abusing the notation  
$\pi(\mathbf{u})=(\pi(u_1),\pi(u_2),\ldots,\pi(u_{n}))$. 
We will denote by $\SLAut(\F_q^{n})$ the group of all semilinear isometries on $\F_q^{n}$. 
Under these definitions, for any $\sigma=(\tau,\pi)\in \SLAut(\F_q^{n})$ 
and $\mathbf{u}\in \F_q^n$,  
there is a monomial matrix $M_{\tau}\in \M(\F_q,n)$ corresponding to $\tau$ such that 
\begin{align}\label{eq.sigma u}
    \sigma(\mathbf{u})=\tau(\pi(\mathbf{u}))=\pi(\mathbf{u})M_{\tau}.     
\end{align}
Note that each monomial matrix $M_{\tau}\in \M(\F_q,n)$ can be decomposed as 
$M_{\tau}=D_{\tau}P_{\tau}$, where $D_{\tau}\in \M(\F_q,n)$ is a diagonal matrix and 
$P_{\tau}\in \M(\F_q,n)$ is a permutation matrix for the permutation $\tau=\left(\begin{array}{ccccc}
    1 & 2 & \cdots & n \\
    \tau_1 & \tau_2 & \cdots & \tau_n
\end{array}\right).$ 
Furthermore, Equation (\ref{eq.sigma u}) can be expressed as $\sigma(\mathbf{u})=\pi(\mathbf{u})D_{\tau}P_{\tau}$. 
Note also that monomial and permutation matrices are non-singular matrices.

For $\sigma=(\tau,\pi)\in \SLAut(\F_q^{n})$ with $\tau$ corresponding to a monomial matrix $M_{\tau}\in \M(\F_q,n)$, 
Carlet $et~al.$ in \cite{sigma_inner_product} introduced the {\em $\sigma$ inner product} 
of $\mathbf{u}$ and $\mathbf{v}\in \F_q^n$ as 
\begin{align*}
    \langle \mathbf{u},\mathbf{v} \rangle_{\sigma}=\sum_{i=1}^{n}u_iv_i',~\text{where}~{{(v_1',v_2',\ldots,v_{n}')=\sigma(\mathbf{v})}}.  
\end{align*} 
Then the {\em $\sigma$ dual} of an $[n,k]_q$ linear code $\C$ is defined by 
\begin{align}
    \C^{\bot_{\sigma}}=\{\mathbf{u}\in \F_q^{n}\mid \langle \mathbf{u},\mathbf{c} \rangle_{\sigma}=0,\ \forall\ \mathbf{c}\in \C \}.     
\end{align}
As usually, $\C$ is called {\em $\sigma$ self-orthogonal} if $\C\subseteq \C^{\bot_{\sigma}}$; 
{\em $\sigma$ dual-containing} if $\C^{\bot_{\sigma}}\subseteq \C$; 
and {\em $\sigma$ self-dual} if $\C=\C^{\bot_{\sigma}}$.

For $0\leq e\leq h-1$, let $\pi_e$ denote the {\em Frobenius automorphism} over $\F_q$ such that 
$\pi_e(\mathbf{u})=(u_1^{p^e}, u_2^{p^e}, \ldots, u_{n}^{p^e})$ and $\pi_e(A)=(a_{ij}^{p^e})$ 
for any $\mathbf{u}=(u_1,u_2,\ldots,u_n) \in \F_q^n$ and $A=(a_{ij})\in \M(\F_q,n)$. 
Also denote $\pi_h=\pi_0$. 
Let ${\tau_ {id}}\in \MAut(\F_q^n)$ be the {\em identity transformation} and  
${\tau_{sym}}\in \MAut(\F_q^{2n})$ be the {\em symplectic transformation}, 
that is, $\tau_ {id}$ corresponds to an identity matrix $I_n$ and 
$\tau_{sym}$ corresponds to a matrix $M_{\tau_{sym}}=\left(\begin{array}{cc}
    O & -I_n \\ I_n & O 
\end{array} \right)$.   
Then it is clear that the $\sigma$ inner product is a generalization of  
the Galois (containing the Euclidean and Hermitian) inner product \cite{FZ2017,sigma_inner_product} and 
the symplectic inner product \cite{XD2021}. 
Specifically speaking, we have the following results (see also \cite{CYW2023,XD2021}). 
\begin{itemize}
    \item [\rm (1)] If $\tau=\tau_ {id}$ and $\pi=\pi_{h-e}$, the $(\tau_ {id}, \pi_{h-e})$ inner product 
    coincides with the {\em $(h-e)$-Galois inner product}. 
    Moreover, if $e=0$, the $(\tau_ {id}, \pi_0)$ inner product is the {\em Euclidean inner product}; 
    and if $e=h/2$ with even $h$, the $(\tau_ {id}, \pi_{h/2})$ inner product 
    is the {\em Hermitian inner product}. 

    \item [\rm (2)] If $\tau=\tau_{sym}$ and $\pi=\pi_0$, the $(\tau_{sym}, \pi_0)$ inner product 
    becomes the {\em symplectic inner product}. 
\end{itemize}

Note that relationships as above also exist for the {\em Euclidean dual code} $\C^{\perp_E}$, 
the {\em Hermitian dual code} $\C^{\perp_H}$, the {\em $e$-Galois dual code} $\C^{\perp_e}$, 
and the {\em symplectic dual code} $\C^{\perp_{S}}$. 
In addition, we also have $\C^{\bot_{\sigma}}=\sigma(\C)^{\bot_E}$ from \cite{sigma_inner_product}, 
where $\sigma(\C)=\pi_e(\C)M_{\tau}=\{\pi_e(\mathbf{c})M_{\tau}\mid \mathbf{c}\in \C\}$.  
According to the following lemma, $\sigma(\C)^{\bot_E}$ can be further determined. 

\begin{lemma}{\rm (\!\!\cite[Lemma 3.2]{CYW2023})}\label{lemma.sigma and Euclidean}
    Let $q=p^h$ be a prime power and $e$ be an integer with $0\leq e\leq h-1$. 
    Let $\C$ be an $[n,k]_{q}$ linear code. 
    If $\sigma=(\tau, \pi_e)\in \SLAut(\F_q^{n})$, 
    where $\tau$ corresponds to a monomial matrix $M_{\tau}\in \M(\F_q,n)$,  
    then $\sigma(\C)^{\perp_E}=\sigma(\C^{\perp_E})(M_{\tau}^TM_{\tau})^{-1}.$
\end{lemma}

\subsection{Toeplitz matrices and non-singular by columns matrices}

In this subsection, we review some basic results on Toeplitz matrices and non-singular by columns (NSC) matrices,  
and then we prove that many NSC matrices can be obtained by known NSC matrices. 

\begin{definition}
    Let $q=p^h$ be a prime power and $A\in \M(\F_q,n)$.
    We say that $A$ is a {\em Toeplitz matrix} if it has constant entries on all diagonals parallel to the main diagonal. 
\end{definition}

\begin{definition}{\rm (\!\! \cite[Definition 3.1]{BN2001})}\label{def.NSC matrix}
    Let $q=p^h$ be a prime power and $A\in \M(\F_q,n)$. 
    Let  $A_{\ell}$ be the $\ell\times n$ matrix consisting of the first $\ell$ rows of $A$  
    and $A_{j_1,j_2,\ldots,j_{\ell}}$ be the $\ell\times \ell$ matrix consisting of columns $j_1,j_2,\ldots,j_{\ell}$ of $A_\ell$.  
    We call $A$ a {\em non-singular by columns (NSC) matrix} 
    if $A_{j_1,j_2,\ldots,j_{\ell}}$ is non-singular for any $1\leq \ell\leq n$ and $1\leq j_1<j_2<\ldots<j_{\ell}\leq n$. 
\end{definition}

The following two lemmas are known.

\begin{lemma}\label{lemma.Toeplitz decompose}{\rm (\!\! \cite[Theorem 2]{SXS2023})}
    Let $q=p^h$ be a prime power and $A\in \M(\F_q,n)$ be any Toeplitz matrix. 
    Then $A=QA^TQ$, where $Q=\adiag(1,1,\dots,1)\in \M(\F_q,n)$ is an anti-diagonal matrix.  
\end{lemma}

\begin{lemma}{\rm (\!\!\cite[Lemma 4.3]{BN2001})}\label{lem.NSC BN2001}
   Let $q=p^h$ be a prime power and $A\in \M(\F_q,n)$ be NSC. Then $Q(A^{-1})^T\in \M(\F_q,n)$ is NSC.
\end{lemma}

\begin{theorem}\label{th.NSC*diagonalmatrix}
    Let $q=p^h$ be a prime power and $A\in \M(\F_q,n)$ be NSC. 
    Then for any diagonal matrix $D=\diag(d_1,d_2,\ldots,d_n)\in \M(\F_q,n)$ with $d_i\neq 0$ for any $1\leq i\leq n$, 
    both $DA$ and $AD$ are NSC.
\end{theorem}
\begin{proof}
    We give a proof for $DA$ and $AD$ can be similarly proved. 
    Since $A$ is NSC and $d_i\neq 0$ for any $1\leq i\leq n$, it holds that  
    $$\det((DA)_{j_1,j_2,\ldots,j_{\ell}})=
      \left(\prod_{i=1}^{\ell}d_{i}\right) \det(A_{j_1,j_2,\ldots,j_{\ell}})\neq 0$$
    for any $1\leq \ell \leq n$ and $1\leq j_1<j_2<\ldots<j_\ell\leq n$. 
    By Definition \ref{def.NSC matrix}, we complete the proof.  
\end{proof}

With above definitions, it is also easy to check that $A\in \M(\F_q,n)$ must be non-singular if $A$ is NSC, 
and $\pi_e(Q)=Q^{-1}=Q^T=Q$ for any $0\leq e\leq h-1$.  
Based on Lemmas \ref{lemma.Toeplitz decompose} and \ref{lem.NSC BN2001}, we then deduce that 
$\pi_e(A)^{-1}Q$ is also NSC for any $0\leq e\leq h-1$ if $A$ is an NSC Toeplitz matrix as follows.

\begin{theorem}\label{th.NSC Galois}
    Let $q=p^h$ be a prime power and $A\in \M(\F_q,n)$ be an NSC Toeplitz matrix. 
    Then for any $0\leq e\leq h-1$,  $\pi_e(A)^{-1}Q\in \M(\F_q,n)$ is NSC. 

\end{theorem}
\begin{proof}
    Since $A$ is NSC, then 
    $$\det(\pi_e(A)_{j_1,j_2,\ldots,j_{\ell}})=\pi_e(\det(A_{j_1,j_2,\ldots,j_{\ell}}))=(\det(A_{j_1,j_2,\ldots,j_{\ell}}))^{p^e}\neq 0$$ 
    for any $1\leq \ell \leq n$, $1\leq j_1<j_2<\ldots<j_\ell\leq n$ and $0\leq e\leq h-1$, 
    which implies that $\pi_e(A)$ is NSC. 
    Since $A$ is a Toeplitz matrix, it follows from Lemma \ref{lemma.Toeplitz decompose} that 
    $$\pi_e(A)^T=\pi_e(QAQ)=Q\pi_e(A) Q.$$ 
    By Lemma \ref{lem.NSC BN2001}, we immediately get that 
    $$Q(\pi_e(A)^{-1})^T=Q(\pi_e(A)^T)^{-1}=Q(Q\pi_e(A)Q)^{-1}=\pi_e(A)^{-1}Q\in \M(\F_q,n)$$ 
    is NSC. 
    This completes the proof. 
\end{proof}

\subsection{Matrix-product codes and their $\sigma$ duals} 

In this subsection, we recall some basic concepts and results on matrix-product codes, 
involving their definition, parameters, and dual codes under both the Euclidean and $\sigma$ inner products. 

\begin{definition}{\rm (\!\! \cite[Definition 2.1]{BN2001})}\label{def.martix product code}
    Let $q=p^h$ be a prime power and $A\in \M(\F_q,s)$. 
    Let $\C_i$ be an $[n,k_i,d_i]_q$ linear code for $i=1,2, \ldots, s$. 
	A {\em matrix-product code} $\C(A)=[\C_1,\C_2,\ldots,\C_s]\cdot A$ 
    is defined as the set of all matrix-products $[\mathbf{c}_1,\mathbf{c}_2,\ldots,\mathbf{c}_s]\cdot A$, 
    where $\mathbf{c}_i=(c_{1i},c_{2i},\ldots,c_{ni})^T\in \C_i$ for $i=1,2,\ldots,s$,  
    $A$ is called the {\em defining matrix} of $\C(A)$, and $\C_1,\C_2,\ldots,\C_s$ are called the {\em input codes} of $\C(A)$. 
A {\rm classical codeword} 
$\mathbf{c}=[\mathbf{c}_1,\mathbf{c}_2,\ldots,\mathbf{c}_s]\cdot A\in \C(A)$ can be expressed as the $n\times s$ matrix 
\begin{align*}
    \mathbf{c}=\left(\begin{array}{cccc}
        \sum_{i=1}^sc_{1i}a_{i1} & \sum_{i=1}^sc_{1i}a_{i2} & \cdots & \sum_{i=1}^sc_{1i}a_{is} \\ 
        \sum_{i=1}^sc_{2i}a_{i1} & \sum_{i=1}^sc_{2i}a_{i2} & \cdots & \sum_{i=1}^sc_{2i}a_{is} \\
        \vdots & \vdots & \ddots & \vdots \\
        \sum_{i=1}^sc_{ni}a_{i1} & \sum_{i=1}^sc_{ni}a_{i2} & \cdots & \sum_{i=1}^sc_{ni}a_{is}
    \end{array}\right).
\end{align*}
Moreover, if we regard $\mathbf{c}_i=(c_{1i}, c_{2i}, \ldots, c_{ni})\in \C_i$ 
as a row vector of length $n$ for $1\leq i\leq s$, 
then any codeword of $\C(A)$ can also be viewed as a row vector of length $sn$, that is,  
$$\mathbf{c}=\left(\sum_{i=1}^sa_{i1}\mathbf{c}_i, \sum_{i=1}^sa_{i2}\mathbf{c}_i, \ldots, \sum_{i=1}^sa_{is}\mathbf{c}_i \right).$$ 
\end{definition}

Note that Definition \ref{def.martix product code} is indeed valid for any $s \times t$ matrix $A$ over $\F_q$. 
However, we only consider the case $s=t$ in this paper. 
The reason for this convention arises from the following series of lemmas.

\begin{lemma}\label{lemma.MP.parameter}{\rm(\!\!\cite{HLR2009,OH2002})}
    Let $q=p^h$ be a prime power and $A\in \M(\F_q,s)$ be non-singular. 
    Let $\C_i$ be an $[n,k_i,d_i]_{q}$ linear code for $i=1,2,\dots,s$. 
    Then $\C(A)$ is an $[sn,\sum_{i=1}^sk_i,\geq d]_{q}$ linear code, 
    where $d=\min\{D_i(A)d_i\mid 1\leq i\leq s\}$ and $D_i(A)$ denotes the minimum distance of the $[s,i]_q$ 
    linear code generated by the first $i$ rows of $A$.  
\end{lemma}

\begin{lemma}{\rm (\!\! \cite[Proposition 6.2]{BN2001})}\label{lem.MP Euclidean}
    Let $q=p^h$ be a prime power and $A\in \M(\F_q,s)$ be non-singular. 
    Let $\C_i$ be an $[n,k_i,d_i]_{q}$ linear code for $i=1,2,\dots,s$. 
    Then the Euclidean dual of $\C(A)$ is 
    \begin{align*}
        \C(A)^{\perp_E}=[\C_1^{\perp_E}, \C_2^{\perp_E}, \ldots, \C_s^{\perp_E}]\cdot (A^{-1})^T.
    \end{align*}
\end{lemma}

\begin{remark}\label{rem.NSC is nice}
    From \cite{BN2001}, the minimum distance of a matrix-product code 
    shown in Lemma \ref{lemma.MP.parameter} may be sharper when the defining matrix is NSC, 
    and in this case one can also obtain a lower bound on the minimum distance of its Euclidean dual code.    
\end{remark}


\begin{lemma}\label{lemma.MP.parameters}{\rm(\!\!\cite[Theorems 3.7 and 6.6]{BN2001})}
    Let $q=p^h$ be a prime power and $A\in \M(\F_q,s)$ be NSC. 
    Let $\C_i$ be an $[n,k_i,d_i]_{q}$ linear code with Euclidean dual distance $d_i^{\perp_E}$ for $i=1,2,\dots,s$. 
    Then $\C(A)$ and its Euclidean dual code 
    have respective parameters 
    $[sn,\sum_{i=1}^sk_i,\geq d ]_{q}~{\rm and}~[sn,sn-\sum_{i=1}^sk_i,\geq d^{\perp_E}]_{q},$ 
    where $d=\min \{(s+1-i)d_i\mid 1\leq i\leq s \}$ and 
    $d^{\perp_E}=\min \{id_i^{\perp_E}\mid 1\leq i\leq s \}$. 
\end{lemma}

Next, we recall some basic results on $\sigma$ duals of matrix-product codes. 
To this end, we need to introduce the Kronecker product of two matrices. 

\begin{definition}{\rm (\!\! \cite[Section 11.4]{K2014})}\label{def.tersor product}
    Let $q=p^h$ be a prime power, $A=(a_{ij})\in \M(\F_q,s)$, and $B\in \M(\F_q,t)$. 
    The {\em Kronecker product} of $A$ and $B$ is defined by 
    \begin{align*}
        A\otimes B = \left(
            \begin{array}{cccc}
                a_{11}B & a_{12}B & \cdots & a_{1s}B \\
                a_{21}B & a_{22}B & \cdots & a_{2s}B \\
                \vdots & \vdots & \ddots & \vdots \\
                a_{s1}B & a_{s2}B & \cdots & a_{ss}B
            \end{array}
        \right)\in \M(\F_q,st). 
    \end{align*}
\end{definition}

\begin{lemma}{\rm (\!\!\cite[Theorem 4.4 and Remark 4.5 (2)]{CYW2023})}\label{lemma.MP.sigma dual}
    Let $q=p^h$ be a prime power and $e$ be an integer with $0\leq e\leq h-1$. 
    Let $\C_i$ be an $[n,k_i,d_i]_{q}$ linear code for $i=1,2,\dots,s$ and $A\in \M(\F_q,s)$ be non-singular. 
    Set $\sigma=(\tau, \pi_e)\in \SLAut(\F_q^{sn})$ and $\sigma'=(\tau', \pi_e)\in \SLAut(\F_q^{n})$, 
    where $\tau$ corresponds to a monomial matrix $M_{\tau}=D_{\tau}P_{\tau}\in \M(\F_q,sn)$ and 
    $\tau'$ corresponds to a monomial matrix $M_{\tau'}=D_{\tau'}P_{\tau'}\in \M(\F_q,n)$. 
    If there is a monomial matrix $M_{\widehat{\tau}}\in \M(\F_q,s)$ such that $M_{\tau}=M_{\widehat{\tau}}\otimes M_{\tau'}$, 
    then the $\sigma$ dual of $\C(A)$ is 
    \begin{align*}
        \C(A)^{\perp_{\sigma}}=[\C_1^{\perp_{\sigma'}}, \C_2^{\perp_{\sigma'}}, \ldots, \C_s^{\perp_{\sigma'}}]\cdot \left(M_{\widehat{\tau}}^T \pi_e(A)^T \right)^{-1}.   
    \end{align*}
\end{lemma}

\begin{lemma}\label{lemma.MP.sigma hull}{\rm (\!\!\cite[Theorem 4.6 and Remark 4.7 (2)]{CYW2023})}
    Let $q=p^h$ be a prime power and $e$ be an integer with $0 \leq e\leq h-1$.  
    Let $\C_i$ be an $[n,k_i,d_i]_{q}$ linear code for $i=1,2,\dots,s$ 
    and $A\in \M(\F_q,s)$. 
    Set $\sigma=(\tau, \pi_e)\in \SLAut(\F_q^{sn})$ and  $\sigma'=(\tau', \pi_e)\in \SLAut(\F_q^{n})$, 
    where $\tau$ corresponds to a monomial matrix $M_{\tau}=D_{\tau}P_{\tau}\in \M(\F_q,sn)$ and 
    $\tau'$ corresponds to a monomial matrix $M_{\tau'}=D_{\tau'}P_{\tau'}\in \M(\F_q,n)$. 
    If there is a monomial matrix $M_{\widehat{\tau}}\in \M(\F_q,s)$ such that $M_{\tau}=M_{\widehat{\tau}}\otimes M_{\tau'}$ 
    and $A {M_{\widehat{\tau}}}^T \pi_e(A)^T=\diag(d_1,d_2,\ldots,d_s)\in \M(\F_q,s)$, 
    then the intersection of $\C(A)$ and $\C(A)^{\perp_{\sigma}}$ is 
    \begin{align}\label{eq.MP.sigma hull}
        \C(A)\cap \C(A)^{\perp_{\sigma}}=[\C_1',\C_2',\ldots,\C_s']\cdot A~{\rm with}~\C_i'=\left\{\begin{array}{ll}
            \C_i, & \mbox{if}~d_i=0, \\
            \C_i\cap \C_i^{\perp_{\sigma'}}, & \mbox{if}~d_i\neq 0.
        \end{array} \right.
    \end{align}
\end{lemma}

\begin{theorem}\label{th.sigma}
    Let the notations and conditions be the same as those in Lemma \ref{lemma.MP.sigma hull}.   
    If $d_i=0$ or $\C_i$ is $\sigma'$ self-orthogonal for $1\leq i\leq s$, 
    then $\C(A)$ is $\sigma$ self-orthogonal. 
\end{theorem}
\begin{proof}
    Note that $\C_i$ is $\sigma'$ self-orthogonal if and only if $\C_i=\C_i\cap \C_i^{\perp_{\sigma'}}$.
    Then combining the given conditions and Equation (\ref{eq.MP.sigma hull}), 
    we have $\C_i'=\C_i$ for any $1\leq i\leq s$. 
    It follows from Lemma \ref{lemma.MP.sigma hull} that $\C(A)\cap \C(A)^{\perp_{\sigma}}=[\C_1,\C_2,\ldots,\C_s]\cdot A=\C(A)$, 
    which implies that $\C(A)\subseteq \C(A)^{\perp_{\sigma}}$.  
    Therefore, $\C(A)$ is $\sigma$ self-orthogonal under the described conditions.  
\end{proof}

\begin{remark}
    Since $\det(M_{\widehat{\tau}})\neq 0$ for any monomial matrix $M_{\widehat{\tau}}\in \M(\F_q,s)$, 
    it follows from the condition $A {M_{\widehat{\tau}}}^T \pi_e(A)^T=\diag(d_1,d_2,\ldots,d_s)$ that 
    $$\det(A)\det(\pi_e(A))=(\det(A))^{p^e+1}=\left(\prod_{i=1}^{s}d_i\right) \det({M_{\widehat{\tau}}})^{-1}\neq 0$$ 
    if and only if $d_i\neq 0$ for any $1\leq i\leq s$. 
    In other words, if $d_i=0$ for some $1\leq i\leq s$, then $\det(A)=0$, and hence, $A$ is not NSC. 
    Combining Remark \ref{rem.NSC is nice}, the case $d_i=0$ for some $1\leq i\leq s$, feasible in Theorem \ref{th.sigma},  
    may not be interesting in practice. 
    As a result, we will expect to find NSC matrices $A$, monomial matrices $M_{\widehat{\tau}}$, 
    and $\sigma'$ self-orthogonal codes $\C_1, \C_2,\ldots, \C_s$  
    that satisfy the conditions required in Theorem \ref{th.sigma} to derive $\sigma$ self-orthogonal matrix-product codes. 
\end{remark}

\section{Four general constructions of $\sigma$ self-orthogonal matrix-product codes associated with Toeplitz matrices}\label{sec.construction}

In this section, we provide four general constructions of $\sigma$ self-orthogonal matrix-product codes.  
The first one relies on the $\sigma'$ dual of a known $\sigma'$ dual-containing matrix-product code; 
the second one is founded on the so-called quasi-$\widehat{\sigma}$ matrix (see Definition \ref{def.quasi-sigma matrix}), 
which also has an application in deriving NSC quasi-unitary matrices;
and the last two ones are based on the utilization of certain special Toeplitz matrices. 
As a byproduct, we also provide a connection between these special Toeplitz matrices 
and the $\widetilde{\tau}$-optimal defining matrices recently introduced in \cite{C2024}.

\subsection{The first general construction via $\sigma'$ duals of known $\sigma'$ dual-containing matrix-product codes} \label{subsec.con1}

We will begin with two important observations.   
\begin{itemize}
    \item [\rm (1)] On one hand, we notice that several families of $q$-ary $\sigma'$ dual-containing matrix-product codes 
    have been constructed in \cite[Section 5]{CYW2023} for some special $\sigma'\in \SLAut(\F_q^{sn})$ 
    with $q\in \{4, 5, 7, 8, 9, 11, 13\}$.  

    \item [\rm (2)] On the other hand, it is {well known} that for any linear code $\C$, 
    $\C^{\perp_E}$ (resp. $\C^{\perp_H}$ and $\C^{\perp_S}$) is Euclidean (resp. Hermitian and symplectic) 
    self-orthogonal if and only if $\C$  is Euclidean (resp. Hermitian and symplectic) dual-containing. 
\end{itemize}
Motivated by the facts above, the following natural question arises: whether 
it is possible to construct $\sigma$ self-orthogonal matrix-product codes by considering the $\sigma'$ dual of 
a known $\sigma'$ dual-containing matrix-product code. 
The following theorem gives an affirmative answer to this question.

\begin{theorem}{\rm (\bf Construction \Rmnum{1})}\label{th.con1}
    Let $q=p^h$ be a prime power and $e,e'$ be two integers with $0\leq e, e'\leq h-1$.   
    Let $\C_i$ be an $[n,k_i,d_i]_{q}$ linear code with Euclidean dual distance $d_i^{\perp_E}$ 
    for $i=1,2,\dots,s$ and $A\in \M(\F_q,s)$ be non-singular. 
    Let $\sigma=(\tau, \pi_e)\in \SLAut(\F_q^{sn})$ and  $\sigma'=(\tau', \pi_{e'})\in \SLAut(\F_q^{sn})$, 
    where $\tau$ corresponds to a monomial matrix $M_{\tau}\in \M(\F_q,sn)$ and 
    $\tau'$ corresponds to a monomial matrix $M_{\tau'}\in \M(\F_q,sn)$. 
    Then the following statements hold. 
    \begin{enumerate}
        \item [\rm (1)] If $e'\equiv h-e~({\rm mod}~h)$ and $\pi_{e}(M_{\tau'})=t M_{\tau}^T$ for some $t\in \F_q^*$, 
        then $\C(A)$ is a $\sigma'$ dual-containing matrix-product code 
        if and only if $\C(A)^{\perp_{\sigma'}}$ is a $\sigma$ self-orthogonal matrix-product code. 
        \item [\rm (2)] If $A$ is NSC, then 
        $\C(A)$ has parameters $[sn,\sum_{i=1}^sk_1,\geq d]_q$ 
        and $\C(A)^{\perp_{\sigma'}}$ has parameters $[sn,sn-\sum_{i=1}^sk_1,\geq d^{\perp_{\sigma'}}]_q$, 
        where $d=\min\left\{(s+1-i)d_{i}\mid 1\leq i\leq s\right\}$ and  
        $d^{\perp_{\sigma'}}=\min \{id_i^{\perp_E}\mid 1\leq i\leq s \}$.  
    \end{enumerate}
\end{theorem}
\begin{proof}
    (1) Recall that $\C^{\perp_{\widetilde{\sigma}}}={\widetilde{\sigma}}(\C)^{\perp_E}$ 
    and ${\widetilde{\sigma}}(\C)=\pi_e(\C)M_{\widetilde{\tau}}$ 
    for any $[n,k]_q$ linear code $\C$ and any ${\widetilde{\sigma}}=(\widetilde{\tau},\pi_e)\in \SLAut(\F_q^n)$, 
    where $\widetilde{\tau}$ corresponds to a monomial matrix $M_{\widetilde{\tau}}\in \M(\F_q,n)$. 
    It then follows from the given conditions, Lemma \ref{lemma.sigma and Euclidean},  
    and the linearity of $\C(A)$ that 
    \begin{align}\label{eq.sigma.sigma.dual}
        \begin{split}
        (\C(A)^{\perp_{\sigma'}})^{\perp_{\sigma}} & = \sigma\left( \left(\C(A)^{\perp_{\sigma'}} \right)^{\perp_E}\right) \left(M_{\tau}^TM_{\tau} \right)^{-1} \\
                                                   & = \sigma\left(\left(\sigma' (\C(A))^{\perp_E}\right)^{\perp_E} \right) M_{\tau}^{-1} \left(M_{\tau}^T \right)^{-1} \\
                                                   & = \pi_{e}\left( \pi_{h-e}(\C(A))M_{\tau'}\right)M_{\tau} M_{\tau}^{-1} \left(M_{\tau}^T \right)^{-1} \\
                                                   & = \C(A) \pi_{e}(M_{\tau'})  \left(M_{\tau}^T \right)^{-1}\\
                                                   & = \C(A). 
        \end{split}
    \end{align}
    Hence, $\C(A)^{\perp_{\sigma'}}\subseteq \C(A)$ if and only if $\C(A)^{\perp_{\sigma'}}\subseteq \left(\C(A)^{\perp_{\sigma'}}\right)^{\perp_{\sigma}}$, 
    that is, $\C(A)$ is $\sigma'$ dual-containing if and only if $\C(A)^{\perp_{\sigma'}}$ is $\sigma$ self-orthogonal. 
    On the other hand, from \cite[Theorem 4.4]{CYW2023}, $\C(A)^{\perp_{\sigma'}}$ is still a matrix-product code. 
    Combining these two aspects, we complete the proof of the result (1). 

    (2) By calculations similar to the proof of the result (1) above, we deduce that 
    $\C(A)^{\perp_{\sigma'}}=\pi_{e'} \left(\C(A)^{\perp_E} \right) \left(M_{\tau'}^T \right)^{-1}$. 
    Hence, $\C(A)^{\perp_{\sigma'}}$ has the same parameters as $\C(A)^{\perp_E}$. 
    Since $A\in \M(\F_q,s)$ is NSC,  
    the parameters of $\C(A)$ and $\C(A)^{\perp_{E}}$ can be immediately 
    obtained from Lemma \ref{lemma.MP.parameters}. 
    This completes the proof the result (2). 
\end{proof}

\begin{example}\label{exam.1}
    Let $q=p^h$ be a prime power and $t\in \F_q^*$ be any element. 
    Recall that several families of $q$-ary $\sigma'$ dual-containing matrix-product codes 
    were constructed in \cite[Section 5]{CYW2023} based on Lemma \ref{lemma.MP.sigma dual}. 
    Assume that $\sigma'=(\tau', \pi_{e'})\in \SLAut(\F_q^{sn})$. 
    One can further note that in \cite[Section 5]{CYW2023}, $\tau'$ always corresponds to a monomial matrix 
    $M_{\tau'}=M_{\widetilde{\tau}} \otimes I_n$ with $M_{\widetilde{\tau}}\in \M(\F_q,s)$ being a monomial matrix. 
    From Theorem \ref{th.con1}, to derive $\sigma$ self-orthogonal matrix-product codes from these $\sigma'$ dual-containing 
    matrix-product codes, it suffices to take $\sigma=(\tau,\pi_{h-e'})\in \SLAut(\F_q^{sn})$ such that $\tau$ corresponds 
    to the monomial matrix 
    \begin{align}\label{eq.exam1}
        \begin{split}
            M_{\tau} & = t^{-1} \pi_{h-e'}(M_{\tau'})^T  \\ 
                      & = \left(t^{-1} I_s \otimes I_n \right) \left(\pi_{h-e'}\left(M_{\widetilde{\tau}}^T \right) \otimes I_n^T\right) \\ 
                      & = \left(t^{-1} \pi_{h-e'}\left(M_{\widetilde{\tau}} \right)^T\right) \otimes I_n. 
        \end{split}
    \end{align}
    Specifically speaking, we have the following results. 

    \begin{enumerate}
        \item [\rm (1)] In Theorems 5.5, 5.8, 5.11, 5.14, 5.17 and 5.20 of \cite{CYW2023}, 
        $e'=0$ and $M_{\widetilde{\tau}}$ being symmetric are restricted. 
        Following Equation (\ref{eq.exam1}), we can take $\sigma=(\tau,\pi_h)=(\tau,\pi_0)$, 
        where $\tau$ corresponds to the monomial matrix 
        $$M_{\tau}=\left(t^{-1}\pi_h\left(M_{\widetilde{\tau}}^T \right)\right) \otimes I_n=\left(t^{-1}\pi_0\left(M_{\widetilde{\tau}}\right)\right) \otimes I_n=t^{-1}M_{\tau'}.$$ 
        Then we immediately get several classes of $q$-ary $\sigma$ self-orthogonal matrix-product codes by taking $\sigma'$ duals of these existing 
        $\sigma'$ dual-containing matrix-product codes.

        \item [\rm (2)] Let $\omega$ be a primitive element of $\F_{4}$.  
        In Theorem 5.2 of \cite{CYW2023}, two classes of quaternary  $\sigma'$ dual-containing matrix-product codes were constructed, 
        where $\sigma'=(\tau', \pi_1)$ and $\tau'$ corresponds to a monomial matrix $M_{\tau'}=M_{\widetilde{\tau}} \otimes I_n$ with $M_{\widetilde{\tau}}$ 
        listed in the second column of Table \ref{tab1}. Following Equation (\ref{eq.exam1}), we can take $\sigma=(\tau,\pi_1)$, 
        where $\tau$ corresponds to a monomial matrix $M_{\tau}$ listed in the third column of Table \ref{tab1}. 
        Again, we directly obtain two classes of quaternary $\sigma$ self-orthogonal matrix-product codes by taking $\sigma'$ duals of the $\sigma'$ dual-containing 
        matrix-product codes constructed in \cite[Theorems 5.2]{CYW2023}.
        \begin{table}[htb!]
            \centering
            \caption{The monomial matrices $M_{\widetilde{\tau}}$ and $M_{\tau}$ in Example \ref{exam.1} (2)}\label{tab1}
            \resizebox{\textwidth}{!}{
            \begin{tabular}{c|c|c|c}
                \hline  
                Finite field $\F_q$ & $M_{\widetilde{\tau}}$ & $M_{\tau}$ & Reference \\ \hline\hline
                $q=4$ & $\left(\begin{array}{cccc}
                    1 & 0 & 0 & 0 \\ 
                    0 & 0 & 0 & \omega+1 \\
                    0 & 0 & 1 & 0 \\
                    0 & \omega & 0 & 0
                \end{array}\right)$      &    $\left(\begin{array}{cccc}
                    t^{-1} & 0 & 0 & 0 \\ 
                    0 & 0 & 0 & t^{-1}\omega^2 \\
                    0 & 0 & t^{-1} & 0 \\
                    0 & t^{-1}\omega & 0 & 0
                \end{array}\right)\otimes I_n$    &   \cite[Theorem 5.2 1)]{CYW2023} \\
                \hline

                $q=4$ & $\left(\begin{array}{cccc}
                    \omega+1 & 0 & 0 & 0 \\ 
                    0 & 0 & 1 & 0 \\
                    0 & \omega & 0 & 0 \\
                    0 & 0 & 0 & \omega+1
                \end{array}\right)$      &    $\left(\begin{array}{cccc}
                    t^{-1}\omega & 0 & 0 & 0 \\ 
                    0 & 0 & t^{-1}\omega^2 & 0 \\
                    0 & t^{-1} & 0 & 0 \\
                    0 & 0 & 0 & t^{-1} \omega
                \end{array}\right)\otimes I_n$    &   \cite[Theorem 5.2 2)]{CYW2023} \\
                \hline
            \end{tabular}}
        \end{table}
    \end{enumerate}
\end{example}

We end this subsection with the following remark.

\begin{remark}\label{rem.th.con1.222}
    Taking into account Equation \eqref{eq.sigma.sigma.dual}, it is easy to conclude that 
    $(\C^{\perp_{\sigma'}})^{\perp_{\sigma}}=\C$ for any $[n,k]_q$ linear code $\C$ 
    provided that $\sigma=(\tau, \pi_e)\in \SLAut(\F_q^{n})$ and  $\sigma'=(\tau', \pi_{e'})\in \SLAut(\F_q^{n})$, 
    where $\tau$ corresponds to a monomial matrix $M_{\tau}\in \M(\F_q,n)$ and 
    $\tau'$ corresponds to a monomial matrix $M_{\tau'}\in \M(\F_q,n)$ 
    satisfying $e'\equiv h-e~({\rm mod}~h)$ and $\pi_{e}(M_{\tau'})=t M_{\tau}^T$ for some $t\in \F_q^*$. 
    In particular, we have the following subcases.  
    \begin{itemize}
        \item [\rm (1)] $\tau=\tau'=\tau_ {id}$. In this case, $M_{\tau}=M_{\tau'}=I_n$, 
        then we have $\pi_{e}(I_n)=t I_n^T$ with $t=1$ for any $0\leq e\leq h-1$. 
        It implies that $(\C^{\perp_{h-e}})^{\perp_{e}}=\C$.  
        Moreover, we get that $(\C^{\perp_E})^{\perp_E}=\C$ and $(\C^{\perp_H})^{\perp_H}=\C$. 

        \item [\rm (2)] $\tau=\tau'=\tau_{sym}$ and $n$ is even. 
        In this case, $M_{\tau}=M_{\tau'}=M_{\tau_{sym}}$, 
        then we have $\pi_{0}(M_{\tau_{sym}})=tM_{\tau_{sym}}^T$ with $t=-1$.  
        It deduces that $(\C^{\perp_{S}})^{\perp_{S}}=\C$. 
    \end{itemize}
    Therefore, according to the proof of Theorem \ref{th.con1}, one has {a unified framework} for the relationship between 
    a linear code and its dual code with respect to a certain inner product. 
    Note also that this framework is not specifically presented in \cite{LL2022sigma,ZKY2023}.
\end{remark}

\subsection{The second general construction via quasi-$\widehat{\sigma}$ matrices obtained from general Toeplitz matrices} \label{subsec.con2}


In this subsection, we present the second general construction of $\sigma$ self-orthogonal matrix-product codes. 
We first introduce the concept of quasi-$\widehat{\sigma}$ matrices, which generalizes 
the well-known quasi-orthogonal matrices and quasi-unitary matrices.   

\begin{definition}\label{def.quasi-sigma matrix}
    Let $q=p^h$ be a prime power and $e$ be an integer with $0\leq e\leq h-1$. 
    Set $A\in \M(\F_q,s)$ and $\widehat{\sigma}=(\widehat{\tau},\pi_e)\in \SLAut(\F_q^{s})$, 
    where $\widehat{\tau}$ corresponds to a monomial matrix $M_{\widehat{\tau}}\in \M(\F_q,s)$. 
    If $A M_{\widehat{\tau}}^T \pi_e(A)^T\in \M(\F_q,s)$ is a diagonal matrix with all nonzero diagonal elements, 
    then we call $A$ a {\em quasi-$\widehat{\sigma}$ matrix}. 
    In particular, a quasi-$\widehat{\sigma}$ matrix coincides with a {\em quasi-orthogonal matrix} 
    if $\widehat{\sigma}=(\tau_ {id}, \pi_0)$; 
    and a {\em quasi-unitary matrix} if $\widehat{\sigma}=(\tau_ {id}, \pi_{h/2})$ and $h$ is even. 
\end{definition}

According to Lemma \ref{lemma.MP.sigma hull} and Theorem \ref{th.sigma}, the construction of quasi-$\widehat{\sigma}$ matrices 
is closely related to the existence of $\sigma$ self-orthogonal matrix-product codes. 
This motivates us to give an explicit way for obtaining NSC quasi-$\widehat{\sigma}$ matrices as follows.

\begin{theorem}
    \label{th.NSC leading principalminors}
    Let $q=p^h$ be a prime power, $r=2e$ if $0\leq e\leq \frac{h}{2}$ and $r=2e-h$ if $\frac{h}{2}< e\leq h-1$. 
    Let $g=\gcd(r,h)$ and $\F_{p^g}$ be the subfield of $\F_{q}$ with $p^g$ elements.
    Suppose that $A\in \M(\F_q,s)$ is an NSC matrix and $M_{\widetilde{\tau}}=D_{\widetilde{\tau}}P_{\widetilde{\tau}}\in \M(\F_{q},s)$ is a monomial matrix 
    such that $AM_{\widetilde{\tau}}\in \M(\F_{p^g},s)$.  
    If all leading principal minors of $AM_{\widetilde{\tau}} \pi_e(AM_{\widetilde{\tau}})^T$ are nonzero, 
    then there is a unit lower triangular matrix $L\in \M(\F_q,s)$  
    such that $LA\in \M(\F_q,s)$ is an NSC quasi-$\widehat{\sigma}$ matrix, 
    where $\widehat{\sigma}=(\widehat{\tau},\pi_e)\in \SLAut(\F_q^s)$ 
    and $\widehat{\tau}$ corresponds to the diagonal matrix $M_{\widehat{\tau}}=D_{\widetilde{\tau}}\pi_e(D_{\widetilde{\tau}})\in \M(\F_q,s)$.
\end{theorem}
\begin{proof}
    Under the given conditions, for any $a\in \F_{p^g}\subseteq \F_q$, 
    we have $\pi_{2e}(a)=\pi_{r}(a)=a$ if $0\leq e\leq \frac{h}{2}$; 
    and $\pi_{2e}(a)=\pi_{r+h}(a)=\pi_{r}(\pi_h(a))=a$ if $\frac{h}{2}< e\leq h-1$. 
    Since $AM_{\widetilde{\tau}}\in \M(\F_{p^g},s)$, we get that 
    $$\pi_e\left(AM_{\widetilde{\tau}} \pi_{e}(AM_{\widetilde{\tau}})^T \right)^T
    =\pi_{2e}(AM_{\widetilde{\tau}}) \pi_e(AM_{\widetilde{\tau}})^T
    =AM_{\widetilde{\tau}} \pi_{e}(AM_{\widetilde{\tau}})^T.$$ 
    It implies that 
    \begin{align}\label{eq.quasi-sigma matrix111}
        AM_{\widetilde{\tau}} \pi_{e}(AM_{\widetilde{\tau}})^T=\left(\begin{array}{cc}
            B_{s-1} & \mathbf{b}_{s-1} \\
            \pi_e({\bf b}_{s-1})^T & b_{s}
        \end{array}
        \right),
    \end{align}
    where $B_{s-1}\in \M(\F_q,s-1)$ satisfies $\pi_e(B_{s-1})^T=B_{s-1}$ and $\det(B_{s-1})\neq 0$, 
    $\mathbf{b}_{s-1}$ is a column vector of length $s-1$ over $\F_{p^g}$, and $\pi_e(b_{s})=b_{s}$. 
    Since $B_{s-1}$ is non-singular, we are able to take 
    \begin{align}\label{eq.quasi-sigma matrix222}
        L_{s-1}=\left(\begin{array}{cc}
            I_{s-1} & \mathbf{0}_{(s-1)\times 1} \\
            -\pi_e({\bf b}_{s-1})^T B_{s-1}^{-1} & 1
        \end{array}
        \right).    
    \end{align}
    Then it can be checked that $L_{s-1}$ is a unit lower triangular matrix and 
    \begin{align}\label{eq.quasi-sigma matrix333}
        \small
        \begin{split}
            &  L_{s-1}AM_{\widetilde{\tau}} \pi_e(M_{\widetilde{\tau}})^T \pi_e(A)^T \pi_e(L_{s-1})^T \\
            = & \left(\begin{array}{cc}
                I_{s-1} & \mathbf{0}_{(s-1)\times 1} \\
                -\pi_e({\bf b}_{s-1})^T B_{s-1}^{-1} & 1
            \end{array}
            \right)\left(\begin{array}{cc}
                B_{s-1} & \mathbf{b}_{s-1} \\
                \pi_e({\bf b}_{s-1})^T & b_{s}
            \end{array}
            \right)\left(\begin{array}{cc}
                I_{s-1} &   -\pi_e(B_{s-1}^{-1})^T \pi_{2e}({\bf b}_{s-1})  \\
                {\bf 0}_{1\times (s-1)} & 1
            \end{array}
            \right) \\ 
            = &\left(\begin{array}{cc}
                B_{s-1} & \mathbf{b}_{s-1} \\
                {\bf 0}_{1\times (s-1)} & -\pi_e({\bf b}_{s-1})^T B_{s-1}^{-1} {\bf b}_{s-1}+b_{s}
            \end{array}
            \right)\left(\begin{array}{cc}
                I_{s-1} &   -(\pi_e(B_{s-1})^T)^{-1} {\bf b}_{s-1}  \\
                {\bf 0}_{1\times (s-1)} & 1
            \end{array}
            \right) \\
            = & \left(\begin{array}{cc}
                B_{s-1} & \mathbf{0}_{(s-1)\times 1} \\
                {\bf 0}_{1\times (s-1)} & -\pi_e({\bf b}_{s-1})^T B_{s-1}^{-1} {\bf b}_{s-1}+b_{s}
            \end{array}
            \right),                  
        \end{split}
    \end{align}
    where $-\pi_e({\bf b}_{s-1})^T B_{s-1}^{-1} {\bf b}_{s-1}+b_{s}\neq 0$ 
    since $$\det\left(L_{s-1}AM_{\widetilde{\tau}} \pi_e(M_{\widetilde{\tau}})^T \pi_e(A)^T \pi_e(L_{s-1})^T\right)
    =\det \left(AM_{\widetilde{\tau}} \pi_{e}(AM_{\widetilde{\tau}})^T \right) \neq 0.$$

    Note that all leading principal minors of $B_{s-1}$ are still nonzero  
    and $\pi_e(B_{s-1})^T=B_{s-1}$, then the above processes shown in Equations 
    (\ref{eq.quasi-sigma matrix111}), (\ref{eq.quasi-sigma matrix222}), and (\ref{eq.quasi-sigma matrix333}) 
    can be repeated for $B_{s-1}$. 
    Note also that they can  be iterated $s-2$ times. 
    Take $L_{s-i-1}'= \left( 
        \begin{array}{cc}
            L_{s-i-1} & O_{(s-i)\times i} \\
            O_{i\times (s-i)} & I_i
        \end{array}
    \right)$ for $1\leq i\leq s-2$ and $L= L_1' L_2' \cdots L_{s-2}' L_{s-1}$. 
    Then it is clear that $L$ is a unit lower triangular matrix such that 
    $$LAM_{\widetilde{\tau}} \pi_e(M_{\widetilde{\tau}})^T \pi_e(A)^T \pi_e(L)^T= (LA) \left(M_{\widetilde{\tau}}\pi_e(M_{\widetilde{\tau}})^T\right) \pi_e(LA)^T=(LA) M_{\widehat{{\tau}}} \pi_e(LA)^T$$ 
    is a diagonal matrix with all nonzero diagonal elements, 
    where $M_{\widehat{{\tau}}}=M_{\widehat{{\tau}}}^T=M_{\widetilde{\tau}}\pi_e(M_{\widetilde{\tau}})^T=D_{\widetilde{\tau}}\pi_e(D_{\widetilde{\tau}})$. 
    Let $\widehat{\sigma}=(\widehat{\tau},\pi_e)$ and $\widehat{\tau}$ 
    corresponds to the diagonal matrix $M_{\widehat{\tau}}$. 
    By Definition \ref{def.quasi-sigma matrix}, 
    $LA$ is just a quasi-$\widehat{\sigma}$ matrix. 
    {Since the matrix $A$ is NSC and $L$ is a unit lower triangular matrix, 
    it follows from the proof of \cite[Theorem 5]{CWC2020} that $LA$ is also NSC.}  
    We have completed the whole proof. 
\end{proof}

\begin{remark}\label{rem.NSC quasi-unitary matrix}
    Let the notations and conditions be the same as those in Theorem \ref{th.NSC leading principalminors}.
    On one hand, if we take $A$ and $M_{\widetilde{\tau}}$ from $\M(\F_{p^g},s)$, 
    then $AM_{\widetilde{\tau}}$ must be in $\M(\F_{p^g},s)$. 
    On the other hand, if we take $M_{\widetilde{\tau}}=I_s$ and $e=h/2$ with even $h$ 
    in Theorem \ref{th.NSC leading principalminors},
    then $M_{\widehat{\tau}}=I_s$, which implies that 
    $\widehat{\sigma}=(\widehat{\tau},\pi_e)=(\tau_ {id},\pi_{h/2})$ coincides with the Hermitian inner product. 
    It means that $LA$ is just an NSC quasi-unitary matrix in this case, which is the same as \cite[Theorem 5]{CWC2020}.
    Consequently, Theorem \ref{th.NSC leading principalminors} can also be regarded as 
    a generalization of \cite[Theorem 5]{CWC2020}. 
\end{remark}

It should be emphasized that Theorem \ref{th.NSC leading principalminors} is important and easy to use. 
Its importance will be shown in Theorem \ref{th.con2} below for constructing $\sigma$ self-orthogonal 
matrix-product codes.  
Combining Theorem \ref{th.NSC leading principalminors} with Theorems \ref{th.NSC*diagonalmatrix}, 
\ref{th.NSC Galois}, and Remark \ref{rem.NSC quasi-unitary matrix}, 
we provide an algorithm for constructing NSC quasi-$\widehat{\sigma}$ matrices in Algorithm \ref{alg.NSC quasi-sigma matrix}. 
In addition, we point out that Algorithm \ref{alg.NSC quasi-sigma matrix} is more efficient compared to the case 
where the matrix $T$ is taken to be other types of matrices in Line $5$ of this algorithm. 
This assertion arises from at least the following two facts.  
\begin{itemize}
    \item [\rm (1)] Toeplitz matrices are easy to store and compute 
    (see \cite{H1996Toeplitz2,M2011Toeplitz3,P2007Toeplitz1,MM1988Toeplitz4} for more details), 
    and many of the submatrices of a Toeplitz matrix are also Toeplitz matrices with smaller sizes. 

    \item [\rm (2)] From Theorems \ref{th.NSC*diagonalmatrix} and \ref{th.NSC Galois}, 
    an NSC Toeplitz matrix is accompanied by three other NSC matrices. 
\end{itemize}

\begin{algorithm}
    \setstretch{0.9}
    \small
    \caption{An algorithm for constructing NSC quasi-$\widehat{\sigma}$ matrices}\label{alg.NSC quasi-sigma matrix}
    \KwIn{A finite field size $q=p^h$, a matrix size $s$, integers $e$, $r$, and $g=\gcd(r,h)$, 
    a monomial matrix $M_{\widetilde{\tau}}=D_{\widetilde{\tau}}P_{\widetilde{\tau}}\in \M(\F_{p^g},s)$,
    and a matrix $Q=\adiag(1,1,\ldots,1)\in \M(\F_{p^g},s)$.}
    \KwOut{An NSC quasi-$\widehat{\sigma}$ matrix in $\M(\F_q,s)$.}
    \tcp{$M_{_{i,i}}$ is the $i\times i$ leading principal submatrix of the matrix $M$.}
    \Begin{
        $A\leftarrow {O}_{s\times s}$. \\
        \While{$A={O}_{s\times s}$}{
           Sample randomly an NSC Toeplitz matrix $T\in \M(\F_{p^g},s)$ and a diagonal matrix $D\in \M(\F_{p^g},s)$ 
           with all nonzero diagonal elements;\\ 
           $\mathcal{PD}_i$ $\leftarrow$ $\left(TDM_{\widetilde{\tau}}\pi_e(TDM_{\widetilde{\tau}})^T\right)_{i,i}$;\\

           $\mathcal{DP}_i$ $\leftarrow$ $\left(DTM_{\widetilde{\tau}}\pi_e(DTM_{\widetilde{\tau}})^T\right)_{i,i}$;\\

           $\mathcal{RD}_i$ $\leftarrow$ $\left(\pi_e(T)^{-1}QDM_{\widetilde{\tau}}\pi_e\left(\pi_e(T)^{-1}QDM_{\widetilde{\tau}}\right)^T\right)_{i,i}$;\\

           $\mathcal{DR}_i$ $\leftarrow$ $\left(D\pi_e(T)^{-1}QM_{\widetilde{\tau}}\pi_e\left(D\pi_e(T)^{-1}QM_{\widetilde{\tau}}\right)^T\right)_{i,i}$;\\

           \uIf {$\det(\mathcal{PD}_i)\neq 0$ for $1\leq i\leq s-1$} {$A\leftarrow TD$;} 
           \uElseIf {$\det(\mathcal{DP}_i)\neq 0$ for $1\leq i\leq s-1$} {$A\leftarrow DT$;}
           \uElseIf {$\det(\mathcal{DP}_i)\neq 0$ for $1\leq i\leq s-1$} {$A\leftarrow \pi_e(T)^{-1}QD$;}
           \ElseIf {$\det(\mathcal{DP}_i)\neq 0$ for $1\leq i\leq s-1$} {$A\leftarrow D\pi_e(T)^{-1}Q$;}
}
           
        $S \leftarrow AM_{\widetilde{\tau}}\pi_e(AM_{\widetilde{\tau}})^T=\left(\begin{array}{cc}
            B_{s-1} & {\bf b}_{s-1} \\ 
            \pi_e({\bf b}_{s-1})^T & b_{s} 
        \end{array} \right)$, where $\pi_e(B_{s-1})^T=B_{s-1}$ is invertible, ${\bf b}_{s-1}\in \F_{p^g}^{s-1}$, and $\pi_e(b_s)=b_{s}$; \\

        $L_{s-1} \leftarrow \left(\begin{array}{cc}
            I_{s-1} & \mathbf{0}_{(s-1)\times 1} \\
            -\pi_e({\bf b}_{s-1})^T B_{s-1}^{-1} & 1
        \end{array}
        \right)$; \\
        \For{$1\leq i\leq s-2$}{
            $B_{s-i}\leftarrow \left(
                \begin{array}{cc}
                    B_{s-i-1} & {\bf b}_{s-i-1} \\ 
                    \pi_e({\bf b}_{s-i-1})^T & b_{s-i}
                \end{array}
            \right)$, where $\pi_e(B_{s-i-1})^T=B_{s-i-1}$ is invertible, ${\bf b}_{s-i-1}\in \F_{p^g}^{s-i-1}$, and $\pi_e(b_{s-i})=b_{s-i}$; \\

            $L_{s-i-1} \leftarrow \left( 
                \begin{array}{cc}
                    I_{s-i-1} & \mathbf{0}_{(s-i-1)\times 1} \\
                    -\pi_e({\bf b}_{s-i-1})^T B_{s-i-1}^{-1} & 1
                \end{array}
            \right);$ \\
            $L_{s-i-1}' \leftarrow \left( 
                \begin{array}{cc}
                    L_{s-i-1} & O_{(s-i)\times i} \\
                    O_{i\times (s-i)} & I_i
                \end{array}
            \right)$;
        }
        $L\leftarrow L_{1}'L_{2}'\cdots L_{s-2}'L_{s-1}$; \\
        $M_{\widehat{\tau}}\leftarrow D_{\widetilde{\tau}}\pi_e(D_{\widetilde{\tau}})$; \\
        $\widehat{\sigma}\leftarrow (\widehat{\tau},\pi_e)$, where $\widehat{\tau}$ corresponds to the matrix $M_{\widehat{\tau}}$; \\
        \Return $LA\in \M(\F_q,s)$ is an NSC quasi-$\widehat{\sigma}$ matrix; 
    }
\end{algorithm}

One has also the following open problem on NSC quasi-unitary matrices, 
which is closely related to the construction of quantum codes as shown in \cite{CWC2020,CC2020QIP,CW2024}.

\begin{problem}{\rm (\!\!\cite[Question 14]{CWC2020})}\label{prob.1}
    Let $q=p^h\neq 2$ be a prime power. For $s=q$ and $q+2\leq s\leq q^2$, 
    do $s\times s$ NSC quasi-unitary matrices exist over $\F_{q^2}$?
\end{problem}

{Some affirmative answers to Problem \ref{prob.1} have been given in \cite[Remark 16]{CWC2020} for the case where $s=q=3$, 
and presented in \cite[Theorem 5.2]{CW2024} for the cases where 
(i) $s=q=3,4,5,7,8,9,11,13,16,17,19$; 
(ii) $s=q+2$ with $q=3,4,5,7,8,9,11,13,16,17,19$; 
(iii) $s=q^{2}-1$ with $q=4,5,7,8,9,11$; 
(iv) $s=q^{2}$ with $q=3,4,5,7,8,9,11,13$.}
Very recently, Cao $et~al.$ in \cite{CW2024} proposed an algorithm for searching NSC quasi-unitary matrices over $\F_{q^2}$ 
by employing the so-called reversely non-singular by columns (RNSC) matrices of type $(V^{-1})^TD$, 
where $V=(x_j^{i-1})_{1\leq i,j\leq s}\in \M(\F_{q^2},s)$ is a non-singular Vandermonde matrix and $D\in \M(\F_{q^2},s)$ 
is a diagonal matrix. For more details on RNSC matrices, we refer to \cite{FLL2014,LL2020}. 
Clearly, $(V^{-1})^TD$ is not a Toeplitz matrix in general. 
Hence, even if we restrict Algorithm \ref{alg.NSC quasi-sigma matrix} to the case of constructing 
NSC quasi-unitary matrices over $\F_{q^2}$, it is different from that given in \cite{CW2024}. 
The following remark gives a further comparison between them.

\begin{remark}
    {Similarly to} \cite[Remark 3.7]{CW2024}, we perform a sampling manner, with the help of the Magma software package \cite{magma}, 
    to count the number of NSC quasi-unitary matrices in $\M(\F_{q^2},s)$ based on NSC Toeplitz matrices by using Algorithm \ref{alg.NSC quasi-sigma matrix} 
    for $(q,s)\in \{(3,3), (3,5), (3,6),  (4,4), (4,6), (4,7),  (5,5), (5,7), (5,8),  (7,7), (7,9), (8,8)\}$. 
    Our sampling results and those of Cao $et~al.$ are listed in the second (resp. third) and fifth (resp. sixth) columns of Table \ref{tab2}, respectively, 
    where the symbol  ``$a/10000$" denotes the number of the NSC quasi-unitary matrices obtained from 
    Algorithm \ref{alg.NSC quasi-sigma matrix} (resp. \cite[Algorithm 1]{CW2024}) after sampling $10000$ times 
    and the symbol  ``$-$" denotes that no sampling results are documented in \cite{CW2024}.  
    From Table \ref{tab2}, it is easy to conclude that our sampling results are 
    more than those presented in \cite[Remark 3.7]{CW2024}. 
    {As we stated before, 
    this provides some advantages due to the use of Toeplitz matrices and it provides an alternative way to obtain NSC quasi-unitary matrices. 
    As an application, one can use these new NSC quasi-unitary matrices to construct quantum codes with good parameters 
    by using the methods documented in \cite{CWC2020,CC2020QIP,CW2024}.}

    \begin{table}[h]
        \centering
        \caption{{Comparison of sampling results between Algorithm \ref{alg.NSC quasi-sigma matrix} and \cite[Algorithm 1]{CW2024}}}\label{tab2}
        \begin{tabular}{c|c|c|c|c|c}
            \hline  
            $(q,s)$ & Algorithm \ref{alg.NSC quasi-sigma matrix} & \cite[Algorithm 1]{CW2024} & $(q,s)$ & Algorithm \ref{alg.NSC quasi-sigma matrix} & \cite[Algorithm 1]{CW2024} \\ \hline\hline
            
            $(3,3)$ &   $8294/10000$ & $5384/10000$ & $(3,5)$  &   $6622/10000$  & $3449/10000$ \\ \hline 

            $(3,6)$ &   $1117/10000$ & $-$  & $(4,4)$ &   $7327/10000$ & $3736/10000$  \\  \hline 

            $(4,6)$ &   $4614/10000$ & $2047/10000$ & $(4,7)$ &   $5673/10000$  & $-$   \\ \hline

            $(5,5)$ &   $7454/10000$ & $4061/10000$ & $(5,7)$ &   $5668/10000$ & $2619/10000$   \\ \hline

            $(5,8)$ &   $4501/10000$ & $-$  & $(7,7)$ &   $7633/10000$  & $3897/10000$ \\ \hline

            $(7,9)$ &   $6235/10000$ & $2913/10000$  & $(8,8)$ &   $7595/10000$  & $3844/10000$ \\ \hline
        \end{tabular}
    \end{table}
\end{remark}

An explicit example on general NSC quasi-$\widehat{\sigma}$ matrix 
is also given below and will be used in the following.

\begin{example}\label{ex.quasi-sigma matrix111}
    Take $p=3$, $h=4$, $e=3$, then $r=2e-h=2$ and $g=\gcd(r,h)=2$. 
    Note that $\F_{3^2}=\{0, 1, \omega^{10}, \omega^{20}, \omega^{30}, 2, \omega^{50}, \omega^{60}, \omega^{70}\} \subseteq \F_{3^4}$, 
    where $\omega$ is a primitive element of $\F_{3^4}$.  
    Let $A$ be a $3\times 3$ NSC Toeplitz matrix over $\F_{3^2}\subseteq \F_{3^4}$ given by  
    $$A=\left(\begin{array}{cccc}
        \omega^{10} & \omega^{50} & \omega^{20} \\ 
        \omega^{30} & \omega^{10} & \omega^{50} \\
        1 & \omega^{30} & \omega^{10}
    \end{array}
    \right)~{\rm and}~M_{\widetilde{\tau}}=\left(\begin{array}{cccc}
        0 & \omega^{10} & 0 \\ 
        2 & 0 & 0 \\
        0 & 0 & \omega^{60}
    \end{array}\right).$$
    One can compute that 
    $$AM_{\widetilde{\tau}}\pi_3(AM_{\widetilde{\tau}})^T=\left(\begin{array}{cccc}
        1 &  0 & 2 \\ 
        0 &  2 & \omega^{30} \\\
        2 & \omega^{10} & 0
    \end{array}
    \right)$$ 
    and all leading principal minors of $AM_{\widetilde{\tau}}\pi_3(AM_{\widetilde{\tau}})^T$ are $1, 2, 2$. 
    Then from the proof of Theorem \ref{th.NSC leading principalminors} 
    (resp. Algorithm \ref{alg.NSC quasi-sigma matrix}), we can take 
    $$L=\left(\begin{array}{cccc}
        1 & 0 & 0 \\
        0 & 1 & 0 \\
        1 & \omega^{10} & 1
    \end{array}
    \right)~{\rm and~hence,}~LA=\left(\begin{array}{cccc}
        \omega^{10} & \omega^{50} & \omega^{20} \\ 
        \omega^{30} & \omega^{10} & \omega^{50} \\
        \omega^{10} & \omega^{60} & \omega^{10}
    \end{array}
    \right)$$ 
    is a $3\times 3$ NSC quasi-$\widehat{\sigma}$ matrix over $\F_{3^4}$, 
    where $\widehat{\sigma}=(\widehat{\tau},\pi_3)$ and $\widehat{\tau}$ corresponds to the diagonal matrix 
    $M_{\widehat{\tau}}=D_{\widetilde{\tau}} \pi_3(D_{\widetilde{\tau}})=\diag(2,1,1)$.
    Moreover, one can check that 
    $(LA)M_{\widehat{\tau}}^T\pi_3(LA)^T=\diag(1,2,1)$. 
    Similarly, more examples can be obtained by applying Theorem \ref{th.NSC leading principalminors} 
    (resp. Algorithm \ref{alg.NSC quasi-sigma matrix}). 
\end{example}

The following result provides our second general construction of $\sigma$ self-orthogonal matrix-product codes.

\begin{theorem}{\rm (\bf Construction \Rmnum{2})}\label{th.con2}
    Let $q=p^h$ be a prime power, $r=2e$ if $0 \leq e\leq \frac{h}{2}$, and $r=2e-h$ if $\frac{h}{2}< e\leq h-1$.  
    Let $g=\gcd(r,h)$ and $\F_{p^g}$ be the subfield field of $\F_{q}$. 
    Let $\C_i$ be an $[n,k_i,d_i]_{q}$ linear code with Euclidean dual distance $d_i^{\perp_E}$ for $i=1,2,\dots,s$ and $A\in \M(\F_q,s)$ be NSC. 
    Set $\sigma=(\tau, \pi_e)\in \SLAut(\F_q^{sn})$ and  $\sigma'=(\tau', \pi_e)\in \SLAut(\F_q^{n})$, 
    where $\tau$ corresponds to a monomial matrix $M_{\tau}=D_{\tau}P_{\tau}\in \M(\F_q,sn)$ and 
    $\tau'$ corresponds to a monomial matrix $M_{\tau'}=D_{\tau'}P_{\tau'}\in \M(\F_q,n)$. 
    If both of the following conditions hold:
    \begin{enumerate}
        \item [\rm (1)] there is a monomial matrix $M_{\widetilde{\tau}}=D_{\widetilde{\tau}}P_{\widetilde{\tau}}\in \M(\F_{q},s)$ such that  
        $AM_{\widetilde{\tau}}^T\in \M(\F_{p^g},s)$ and all principal minors of $AM_{\widetilde{\tau}}^T \pi_e(AM_{\widetilde{\tau}})^T$ are nonzero; 
        \item [\rm (2)] $M_{\tau}=(D_{\widetilde{\tau}}\pi_e(D_{\widetilde{\tau}}))\otimes M_{\tau'}$ and $\C_i$ is $\sigma'$ self-orthogonal for any $1\leq i\leq s$, 
    \end{enumerate}
    then there is a unit lower triangular matrix $L\in \M(\F_q, s)$ such that 
    $\C(LA)$ is an $[sn,\sum_{i=1}^{s}k_i,\geq d]_q$ $\sigma$ self-orthogonal matrix-product code, where $d=\min\{(s+1-i)d_i\mid 1\leq i\leq s\}$. 
    Moreover, $\C(LA)^{\perp_{\sigma}}$ has parameters $[sn,sn-\sum_{i=1}^{s}k_i,\geq d^{\perp_{\sigma}}]_q$, 
    where $d^{\perp_{\sigma}}=\min  \{id_i^{\perp_E}\mid 1\leq i\leq s  \}$. 
\end{theorem}
\begin{proof}
    Denote $M_{\widehat{\tau}}=D_{\widetilde{\tau}}\pi_e(D_{\widetilde{\tau}})$, then $M_{\widehat{\tau}}^T=M_{\widehat{\tau}}\in \M(\F_q,s)$ is a monomial matrix. 
    It follows from the condition (1) and Theorem \ref{th.NSC leading principalminors} that 
    there is a unit lower triangular matrix $L\in \M(\F_q,s)$ such that $LA$ is NSC and 
    $$(LA) M_{\widehat{\tau}}^T  \pi_e(LA)^T=(LA) (D_{\widetilde{\tau}}\pi_e(D_{\widetilde{\tau}})) \pi_e(LA)^T=\diag(d_1,d_2,\ldots,d_s)\in \M(\F_q,s)$$ 
    with $d_i\neq 0$ for any $1\leq i\leq s$. 
    Combining the condition (2) and Theorem \ref{th.sigma}, 
    we directly conclude that $\C(LA)$ is a $\sigma$ self-orthogonal matrix-product code.  
    By an argument similar to that in the proof of Theorem \ref{th.con1} (2), 
    $\C(LA)$ and $\C(LA)^{\perp_{\sigma}}$ have the described parameters. 
\end{proof}

\begin{example}\label{example.con2}
    According to \cite[Theorem 4.5]{QCWL2024}, there exist $(\tau_ {id},\pi_3)$ ($i.e.,$ $1$-Galois) self-orthogonal $[81,k,82-k]_{3^4}$ 
    MDS codes for any $1\leq k\leq 20$. 
    Take $\sigma=(\tau,\pi_3)$ and $\sigma'=(\tau_ {id},\pi_3)$, where $\tau$ corresponds to the monomial matrix 
    $M_{\tau}=M_{\widehat{\tau}}\otimes I_{81}$ with $M_{\widehat{\tau}}=\diag(2,1,1)$ as in Example \ref{ex.quasi-sigma matrix111}. 
    Put $\C_i=[81,k_i,82-k_i]_{3^4}$ for $1\leq k_i\leq 20$ and $1\leq i\leq 3$. 
    Combining Example \ref{ex.quasi-sigma matrix111} with Theorem \ref{th.con2}, 
    we immediately get $\sigma$ self-orthogonal matrix-product codes with parameters 
    $$\left[243,k_1+k_2+k_3,\geq \min\{(4-i)(82-k_i)\mid 1\leq i\leq 3\}\right]_{3^4}.$$
    We list some explicit parameters of them in Table \ref{tab3}, where $\C_1=\C_2=[81,20,62]_{3^4}$ is taken.

    \begin{table}[h]
        \centering
        \caption{Some $\sigma$ self-orthogonal matrix-product codes derived from Theorem \ref{th.con2}}\label{tab3}
        \resizebox{\textwidth}{!}{
        \begin{tabular}{c|c|c|c}
            \hline  
            $\C_3$ & $\sigma$ self-orthogonal matrix-product code & $\C_3$ & $\sigma$ self-orthogonal matrix-product code \\ \hline\hline
            
            $[81,1,81]_{3^4}$ & $[243,41,\geq 81]_{3^4}$ & $[81,2,80]_{3^4}$ & $[243,42,\geq 80]_{3^4}$ \\ \hline 
            
            $[81,3,79]_{3^4}$ & $[243,43,\geq 79]_{3^4}$ & $[81,4,78]_{3^4}$ & $[243,44,\geq 76]_{3^4}$  \\  \hline 

            $[81,5,77]_{3^4}$ & $[243,45,\geq 77]_{3^4}$ & $[81,6,76]_{3^4}$ & $[243,46,\geq 76]_{3^4}$ \\ \hline

            $[81,7,75]_{3^4}$ & $[243,47,\geq 75]_{3^4}$ & $[81,8,74]_{3^4}$ & $[243,48,\geq 74]_{3^4}$   \\ \hline

            $[81,9,73]_{3^4}$ & $[243,49,\geq 73]_{3^4}$ & $[81,10,72]_{3^4}$ & $[243,50,\geq 72]_{3^4}$ \\ \hline

            $[81,11,71]_{3^4}$ & $[243,51,\geq 71]_{3^4}$ & $[81,12,70]_{3^4}$ & $[243,52,\geq 70]_{3^4}$ \\ \hline

            $[81,13,69]_{3^4}$ & $[243,53,\geq 69]_{3^4}$ & $[81,14,68]_{3^4}$ & $[243,54,\geq 68]_{3^4}$ \\ \hline

            $[81,15,67]_{3^4}$ & $[243,55,\geq 67]_{3^4}$ & $[81,16,66]_{3^4}$ & $[243,56,\geq 66]_{3^4}$ \\ \hline

            $[81,17,65]_{3^4}$ & $[243,57,\geq 65]_{3^4}$ & $[81,18,64]_{3^4}$ & $[243,58,\geq 64]_{3^4}$ \\ \hline

            $[81,19,63]_{3^4}$ & $[243,59,\geq 63]_{3^4}$ & $[81,20,62]_{3^4}$ & $[243,60,\geq 62]_{3^4}$ \\ \hline

        \end{tabular}}
    \end{table}
\end{example}

\subsection{The third and fourth general constructions via special Toeplitz matrices}

In this subsection, we give our last two general constructions of $\sigma$ self-orthogonal matrix-product codes 
such that their defining matrices are Toeplitz matrices. 
As a byproduct of these two constructions, we also find an interesting application of these Toeplitz matrices 
in the so-called $\widetilde{\tau}$-optimal defining matrices. 
The following result provides the third general construction.

\begin{theorem}{\rm (\bf Construction \Rmnum{3})}\label{th.con3}
    Let $q=p^h$ be a prime power and $e$ be an integer with $0\leq e\leq h-1$.  
    Let $\C_i$ be an $[n,k_i,d_i]_{q}$ linear code with Euclidean dual distance $d_i^{\perp_E}$ for $i=1,2,\dots,s$ and $A\in \M(\F_q,s)$ be an NSC Toeplitz matrix. 
    Set $\sigma=(\tau, \pi_e)\in \SLAut(\F_q^{sn})$ and  $\sigma'=(\tau', \pi_e)\in \SLAut(\F_q^{n})$, 
    where $\tau$ corresponds to a monomial matrix $M_{\tau}=D_{\tau}P_{\tau}\in \M(\F_q,sn)$ and 
    $\tau'$ corresponds to a monomial matrix $M_{\tau'}=D_{\tau'}P_{\tau'}\in \M(\F_q,n)$. 
    If both of the following conditions hold:
    \begin{enumerate}
        \item [\rm (1)] there is a monomial matrix $M\in \M(\F_q,s)$ such that $\pi_e(A)MA=DQ$ for some diagonal matrix $D\in \M(\F_q,s)$; 
        \item [\rm (2)] $M_{\tau}=(MQ)\otimes M_{\tau'}$ and $\C_i$ is $\sigma'$ self-orthogonal for any $1\leq i\leq s$, 
    \end{enumerate}
    then $\C(A)$ is an $[sn,\sum_{i=1}^{s}k_i,\geq d]_q$ $\sigma$ self-orthogonal matrix-product code, where $d=\min\{(s+1-i)d_i\mid 1\leq i\leq s\}$. 
    Moreover,  $\C(A)^{\perp_{\sigma}}$ has parameters $[sn,sn-\sum_{i=1}^{s}k_i,\geq d^{\perp_{\sigma}}]_q$, 
    where $d^{\perp_{\sigma}}=\min  \{id_i^{\perp_E}\mid 1\leq i\leq s  \}$. 
\end{theorem}
\begin{proof}
    Denote $M_{\widehat{\tau}}=MQ$. 
    Since $M\in \M(\F_q,s)$ is a monomial matrix and $Q=\adiag(1,1,\ldots,1)\in \M(\F_q,s)$, 
    it is easy to see that $M_{\widehat{\tau}}\in \M(\F_q,s)$ is still a monomial matrix. 
    Noting that $\pi_e(A)MA=DQ$ and $A$ is NSC, we further get that all diagonal elements of $D$ are nonzero 
    since $$\det(D)=\det(\pi_e(A))\det(M)\det(A)\det(Q)^{-1}\neq 0.$$ 
    It then follows from the condition (1) and Lemma \ref{lemma.Toeplitz decompose} that 
    $$AM_{\widehat{\tau}}^T\pi_e(A)^T = \left(\pi_e(A)M_{\widehat{\tau}}A^T \right)^T= \left(\pi_e(A)M(QA^T) \right)^T= (\pi_e(A)M(AQ))^T= D.$$
    Combining the condition (2) and Theorem \ref{th.sigma},  
    it is straightforward to deduce that $\C(A)$ is a $\sigma$ self-orthogonal matrix-product code. 
    Again, the parameters of $\C(A)$ and $\C(A)^{\perp_{\sigma}}$ can be obtained by a similar argument as the one in the proof of Theorem \ref{th.con1} (2).        
\end{proof}

\begin{example}\label{exam.con3}
    Let $q=2^6$ and $e=0$.   
    Let $A\in \M(\F_{2^6},3)$ be an NSC Toeplitz matrix, $M\in \M(\F_{2^6},3)$ be a monomial matrix, 
    and $D=\diag(\omega^{36}, \omega^{54},1)\in \M(\F_{2^6},3)$ be a diagonal matrix given by  
    $$A=\left(\begin{array}{ccc}
        1 & \omega^{54} & \omega^{27} \\
        \omega^{36} & 1 & \omega^{54} \\ 
        \omega^{54} & \omega^{36} & 1  
    \end{array}
    \right)~{\rm and}~M=\left(\begin{array}{ccc}
        \omega^{27} & 0 & 0 \\
        0 & \omega^{54} & 0 \\ 
        0 & 0 & \omega^{27}
    \end{array}
    \right),$$
    where $\omega$ is a primitive element of $\F_{2^6}$. 
    It can be checked that $\pi_0(A)MA=DQ$.  
    According to \cite[Theorem 3]{LCC2018}, there exist $(\tau_ {id},\pi_0)$ ($i.e.,$ Euclidean) self-orthogonal 
    $[64,k,65-k]_{2^6}$ MDS codes for any $1\leq k\leq 32$. 
    Take $\sigma=(\tau,\pi_0)$ and $\sigma'=(\tau_ {id},\pi_0)$, where $\tau$ corresponds to the monomial matrix 
    $M_{\tau}=(MQ)\otimes I_{64}=\adiag(\omega^{27},\omega^{54},\omega^{27})\otimes I_{64}$. 
    Let  $\C_i=[64,k_i,65-k_i]_{2^6}$ for $1\leq k_i\leq 32$ and $1\leq i\leq 3$. 
    With Theorem \ref{th.con3}, we immediately get $\sigma$ self-orthogonal matrix-product codes with parameters 
    $$\left[192,k_1+k_2+k_3,\geq \min\{(4-i)(65-k_i)\mid 1\leq i\leq 3\}\right]_{2^6}.$$
    For more clarity, we list some explicit parameters of them in Table \ref{tab4}, where $\C_1=\C_2=[64,32,33]_{2^6}$ is taken.

    \begin{table}[h]
        \centering
        \caption{Some $\sigma$ self-orthogonal matrix-product codes derived from Theorem \ref{th.con3}}\label{tab4}
        \resizebox{\textwidth}{!}{
        \begin{tabular}{c|c|c|c}
            \hline  
            $\C_3$ & $\sigma$ self-orthogonal matrix-product code & $\C_3$ & $\sigma$ self-orthogonal matrix-product code \\ \hline\hline
            
            $[64,1,64]_{2^6}$ & $[192,65,\geq 64]_{2^6}$ & $[64,2,63]_{2^6}$ & $[192,66,\geq 63]_{2^6}$ \\ \hline  

            $[64,3,62]_{2^6}$ & $[192,67,\geq 62]_{2^6}$ & $[64,4,61]_{2^6}$ & $[192,68,\geq 61]_{2^6}$  \\  \hline

            $[64,5,60]_{2^6}$ & $[192,69,\geq 60]_{2^6}$ & $[64,6,59]_{2^6}$ & $[192,70,\geq 59]_{2^6}$ \\ \hline

            $[64,7,58]_{2^6}$ & $[192,71,\geq 58]_{2^6}$ & $[64,8,57]_{2^6}$ & $[192,72,\geq 57]_{2^6}$   \\ \hline

            $[64,9,56]_{2^6}$ & $[192,73,\geq 56]_{2^6}$ & $[64,10,55]_{2^6}$ & $[192,74,\geq 55]_{2^6}$ \\ \hline

            $[64,11,54]_{2^6}$ & $[192,75,\geq 54]_{2^6}$ & $[64,12,53]_{2^6}$ & $[192,76,\geq 53]_{2^6}$ \\ \hline

            $[64,13,52]_{2^6}$ & $[192,77,\geq 52]_{2^6}$ & $[64,14,51]_{2^6}$ & $[192,78,\geq 51]_{2^6}$ \\ \hline

            $[64,15,50]_{2^6}$ & $[192,79,\geq 50]_{2^6}$ & $[64,16,49]_{2^6}$ & $[192,80,\geq 49]_{2^6}$ \\ \hline

            $[64,17,48]_{2^6}$ & $[192,81,\geq 48]_{2^6}$ & $[64,18,47]_{2^6}$ & $[192,82,\geq 47]_{2^6}$ \\ \hline

            $[64,19,46]_{2^6}$ & $[192,83,\geq 46]_{2^6}$ & $[64,20,45]_{2^6}$ & $[192,84,\geq 45]_{2^6}$ \\ \hline

            $[64,21,44]_{2^6}$ & $[192,85,\geq 44]_{2^6}$ & $[64,22,43]_{2^6}$ & $[192,86,\geq 43]_{2^6}$ \\ \hline

            $[64,23,42]_{2^6}$ & $[192,87,\geq 42]_{2^6}$ & $[64,24,41]_{2^6}$ & $[192,88,\geq 41]_{2^6}$ \\ \hline

            $[64,25,40]_{2^6}$ & $[192,89,\geq 40]_{2^6}$ & $[64,26,39]_{2^6}$ & $[192,90,\geq 39]_{2^6}$ \\ \hline

            $[64,27,38]_{2^6}$ & $[192,91,\geq 38]_{2^6}$ & $[64,28,37]_{2^6}$ & $[192,92,\geq 37]_{2^6}$ \\ \hline

            $[64,29,36]_{2^6}$ & $[192,93,\geq 36]_{2^6}$ & $[64,30,35]_{2^6}$ & $[192,94,\geq 35]_{2^6}$ \\ \hline

            $[64,31,34]_{2^6}$ & $[192,95,\geq 34]_{2^6}$ & $[64,32,33]_{2^6}$ & $[192,96,\geq 33]_{2^6}$ \\ \hline
        \end{tabular}}
    \end{table}
\end{example}

If we add some extra restrictions to the Toeplitz matrices $A$ used in Theorem \ref{th.con3} ({\bf Construction \Rmnum{3}}), 
we can obtain the fourth general construction of $\sigma$ self-orthogonal matrix-product codes as follows, 
in which $\C_1, \C_2,\ldots, \C_s$ are no longer required to be $\sigma'$ self-orthogonal.

\begin{theorem}{\rm (\bf Construction \Rmnum{4})}\label{th.con4}
    Let $q=p^h$ be a prime power and $e$ be an integer with $0\leq e\leq h-1$. 
    Let $\C_i$ be an $[n,k_i,d_i]_{q}$ linear code with Euclidean dual distance $d_i^{\perp_E}$ for each $i=1,2,\dots,s$ and $A\in \M(\F_q,s)$.  
    Set $\sigma=(\tau, \pi_e)\in \SLAut(\F_q^{sn})$ and $\sigma'=(\tau', \pi_e)\in \SLAut(\F_q^{n})$, 
    where $\tau$ corresponds to a monomial matrix $M_{\tau}\in \M(\F_q,sn)$ and 
    $\tau'$ corresponds to a monomial matrix $M_{\tau'}\in \M(\F_q,n)$ such that $M_{\tau}=D\otimes M_{\tau'}$ 
    for some diagonal matrix $D\in \M(\F_q,s)$. 
    Then the following statements hold. 
    \begin{enumerate}
        \item [\rm (1)] If $A$ is a non-singular Toeplitz matrix,  then  
        $\C(A)^{\perp_{\sigma}}=[\C_s^{\perp_{\sigma'}}, \C_{s-1}^{\perp_{\sigma'}}, \ldots, \C_1^{\perp_{\sigma'}}]\cdot \pi_e(A)^{-1}QD^{-1}$.
        
        \item [\rm (2)] If $A$ is a non-singular Toeplitz matrix and $\C_{i} \subseteq \C_{s+1-i}^{\perp_{\sigma'}}$ for $1\leq i\leq s$, 
        then $\C(\pi_e(A)^{-1}QD^{-1}) \subseteq \C(A)^{\perp_{\sigma}}.$ 
        Moreover, if $\pi_e(A)A=QD^{-1}$, then $\C(A)$ is a $\sigma$ self-orthogonal matrix-product code.
        
        \item [\rm (3)] If $A$ is an NSC Toeplitz matrix, then $\C(\pi_e(A)^{-1}QD^{-1})$ has parameters $[sn,\sum_{i=1}^{s}k_i,\geq d]_q$ 
        and $\C(A)^{\perp_{\sigma}}$ has parameters $[sn,sn-\sum_{i=1}^{s}k_i,\geq d^{\perp_{\sigma}}]_q$, 
        where $d=\min \left\{(s+1-i)d_{i}\mid 1\leq i\leq s \right\}$ and $d^{\perp_{\sigma}}=\min  \{id_i^{\perp_E}\mid 1\leq i\leq s  \}$.
   
    \end{enumerate}
\end{theorem}
\begin{proof}
    (1) 
    Recall that $\pi_e(Q)=Q^T=Q^{-1}=Q$ for $0\leq e\leq h-1$. 
    Regard $Q$ as the permutation matrix corresponding to the reverse permutation 
    $\left(\begin{array}{cccc}
        1 & 2 & \cdots & s \\ 
        s & s-1 & \cdots & 1
    \end{array}\right).$ 
    Since $M_{\tau}=D\otimes M_{\tau'}$ and both $M_{\tau}$ and $M_{\tau'}$ are monomial matrices, 
    we easily know that each diagonal element of $D$ is nonzero. 
    It then follows form Lemma \ref{lemma.MP.sigma dual} that  
    \begin{align*}
        \begin{split}
            \C(A)^{\bot_\sigma} & = [\C_1^{\bot_{\sigma'}},\C_2^{\bot_{\sigma'}},\dots,\C_s^{\bot_{\sigma'}}]\cdot (D^T \pi_e(A)^T)^{-1} \\
                                & = [\C_1^{\bot_{\sigma'}},\C_2^{\bot_{\sigma'}},\dots,\C_s^{\bot_{\sigma'}}]\cdot (Q\pi_e(A) Q)^{-1}D^{-1} \\
                                & = [\C_1^{\bot_{\sigma'}},\C_2^{\bot_{\sigma'}},\dots,\C_s^{\bot_{\sigma'}}]\cdot Q\pi_e(A)^{-1} Q D^{-1}\\
                                & = [\C_s^{\bot_{\sigma'}},\C_{s-1}^{\bot_{\sigma'}},\dots,\C_1^{\bot_{\sigma'}}]\cdot \pi_e(A)^{-1} Q D^{-1}.
        \end{split}
    \end{align*}
    This completes the proof of the result (1).  

    (2) Note that $\C(\pi_e(A)^{-1}QD^{-1})=[\C_1,\C_2,\ldots,\C_s]\cdot \pi_e(A)^{-1}QD^{-1}$. 
    Combining the result (1) above and the given conditions, the result (2) clearly holds. 

    (3) Since $A$ is an NSC Toeplitz matrix and $D\in \M(\F_q,s)$ is a diagonal matrix, 
    it deduces from Theorems \ref{th.NSC*diagonalmatrix} and \ref{th.NSC Galois} that $\pi_e(A)^{-1}Q D^{-1}\in \M(\F_q,s)$ is also an NSC matrix. 
    With an argument similar to that in the proof of Theorem \ref{th.con1} (2), the result (3) follows directly.  
\end{proof}

{
In \cite{C2024}, Cao introduced the so-called $\widetilde{\tau}$-optimal defining ($\widetilde{\tau}$-OD) matrices, 
which can be taken as defining matrices of matrix-product codes and are closely related to the constructions 
of quantum codes with good parameters.}
As a byproduct, we get a connection between Toeplitz matrices and 
the $\widetilde{\tau}$-OD matrices when $D=I_s$ and $e=0$ are fixed 
in the condition (1) of Theorem \ref{th.con3} (resp. Theorem \ref{th.con4} (2)). 
To this end, we first recall the definition of $\widetilde{\tau}$-OD matrices, 
and from it one realizes that $\widetilde{\tau}$-OD matrices 
are indeed a generalization of NSC quasi-orthogonal matrices mentioned in Definition \ref{def.quasi-sigma matrix}. 

\begin{definition}{\rm (\!\! \cite[Definition 5.3]{C2024})}\label{def.tau-OD}
    Let $q=p^h$ be a prime power and $A\in \M(\F_q,s)$ be NSC. 
    We call $A$ a {\em $\widetilde{\tau}$-optimal defining ($\widetilde{\tau}$-OD) matrix} 
    if $AA^T=DP_{\widetilde{\tau}}$ for some diagonal matrix $D\in \M(\F_q,s)$ and 
    permutation matrix $P_{\widetilde{\tau}}\in \M(\F_q,s)$ corresponding to the 
    permutation $\widetilde{\tau}$. 
\end{definition}

{The following remark gives a connection between Toeplitz matrices and $\widetilde{\tau}$-OD matrices.}

\begin{remark}{\bf (A connection between Toeplitz matrices and $\widetilde{\tau}$-OD matrices)}\label{rem.tau-OD}
    Suppose that $A\in \M(\F_q,s)$ is an NSC Toeplitz matrix satisfying $\pi_e(A)MA=Q$ 
    for some monomial matrix $M\in \M(\F_q,s)$.  We have the following facts. 
    \begin{enumerate}
        \item [\rm (1)] From Lemma \ref{lemma.Toeplitz decompose}, we conclude that 
        \begin{align*}
            A \pi_e(A)^T  = A \pi_e(QAQ) = A Q (QA^{-1}M^{-1}) Q = M^{-1} Q \in \M(\F_q,s) 
        \end{align*}
        is a monomial matrix. Since $M^{-1}\in M(\F_q,s)$ is also monomial, 
        then we can fix  that $M^{-1}=D_{\tau'}P_{\tau'}$, where $D_{\tau'}\in \M(\F_q,s)$ is a diagonal matrix 
        and $P_{\tau'}\in \M(\F_q,s)$ is a permutation matrix corresponding to the permutation 
        $\tau'=\left(\begin{array}{cccc}
            1 & 2 & \cdots & s \\ 
            \tau'_1 & \tau'_2 & \cdots & \tau'_s
        \end{array}\right)$. 
        It follows that $P_{\tau'}Q\in \M(\F_q,s)$ is a permutation matrix corresponding to the permutation 
        $$\widetilde{\tau}=\left(\begin{array}{cccc}
            1 & 2 & \cdots & s \\ 
            s+1-\tau'_1 & s+1-\tau'_2 & \cdots & s+1-\tau'_s
        \end{array}\right).$$ 
        By Definition \ref{def.tau-OD}, the NSC Toeplitz matrix $A$ is just a $\widetilde{\tau}$-OD matrix in the case of $e=0$. 

        \item [\rm (2)] {On the other hand, we note that for $s\in \{3,4\}$, 
        the matrices of the types $VD$, $LV$ and $LVD$ 
        have been used to find $\widetilde{\tau}$-OD matrices in \cite[Section 5]{C2024}, where 
        $V=(x_j^{i-1})_{1\leq i,j\leq s}$ is a non-singular Vandermonde matrix, 
        $D$ is a diagonal matrix and $L$ is a lower triangular matrix.} 
        It is also easily checked that {$VD$, $LV$ and $LVD$} are not Toeplitz matrices in general.  
        For some specific examples, one can see \cite[Section 5]{C2024}.

        \item [\rm (3)] {Combining (1) and (2) above, we conclude that the Toeplitz matrices $A$ used in 
        Theorem \ref{th.con3} ({\bf Construction \Rmnum{3}}) 
        (resp. Theorem \ref{th.con4} ({\bf Construction \Rmnum{4}}) (2))
        are able to find new $\widetilde{\tau}$-OD matrices of size $s\times s$, where $s$ is not restricted to $3$ or $4$.  
        Therefore, more quantum codes with good parameters can be further obtained 
        by using these new $\widetilde{\tau}$-OD matrices and the methods documented in \cite{C2024}.}

        \item [\rm (4)] Furthermore, since we restrict $e = 0$ and $D = I_s$ when considering $\widetilde{\tau}$-OD matrices, 
        then we can also regard the Toeplitz matrices $A$ used in Theorem \ref{th.con3} ({\bf Construction \Rmnum{3}}) 
        (resp. Theorem \ref{th.con4} ({\bf Construction \Rmnum{4}}) (2))
        as generalizations of $\widetilde{\tau}$-OD matrices. 
        In other words, if the restrictions $e = 0$ and $D = I_s$ are relaxed, 
        the Toeplitz matrices $A$ used in Theorem \ref{th.con3} ({\bf Construction \Rmnum{3}}) 
        (resp. Theorem \ref{th.con4} ({\bf Construction \Rmnum{4}}) (2)) 
        may not be $\widetilde{\tau}$-OD matrices for any $\widetilde{\tau}$.     
    \end{enumerate} 
\end{remark}

To illustrate the facts in Remark \ref{rem.tau-OD} we provide the following two examples.

\begin{example}\label{exam.tau-OD111}
    Let $A\in \M(\F_{2^3},3)$ be an NSC Toeplitz matrix and $M\in \M(\F_{2^3},3)$ be a monomial matrix given by 
    $$A=\left(\begin{array}{ccc}
        1 & \omega^2 & \omega^3 \\
        \omega^3 & 1 & \omega^2 \\ 
        \omega^2 & \omega^3 & 1  
    \end{array}
    \right)~{\rm and}~M=\left(\begin{array}{ccc}
        0 & 0 & \omega^6 \\
        0 & \omega^6 & 0 \\ 
        \omega^6 & 0 & 0
    \end{array}
    \right),$$
    where $\omega$ is a primitive element of $\F_{2^3}$. 
    It can be easily checked that 
    $$\pi_0(A)MA=Q ~ {\rm and} ~ M^{-1}=\diag(\omega,\omega,\omega)Q$$ is a monomial matrix. 
    From Remark \ref{rem.tau-OD} (1), we immediately get that $A$ is a $\tau_ {id}$-OD matrix. 
\end{example}

\begin{example}\label{exam.tau-OD222}  
    Let $A\in \M(\F_{2^6},3)$ be the NSC Toeplitz matrix shown in Example \ref{exam.con3}. 
    It is easy to check that 
    $$AA^T=\left(\begin{array}{ccc}
        \omega^{27} & \omega^{45} & \omega^{54} \\
        \omega^{45} & \omega^{18} & 0 \\ 
        \omega^{54} & 0 & \omega^{18}  
    \end{array}
    \right)$$ 
    is not a $\widetilde{\tau}$-OD matrix for any permutation $\widetilde{\tau}$, 
    where  $\omega$ is a primitive element of $\F_{2^6}$. 
    Combining Example \ref{exam.con3}, we conclude that $A$ is indeed a candidate 
    for Theorem \ref{th.con3} ({\bf Construction \Rmnum{3}}) but not for \cite{C2024}. 
\end{example}

\subsection{Comparisons of the four general constructions}

In this subsection, we give comparisons among the four general constructions of 
$\sigma$ self-orthogonal matrix-product codes proposed in above subsections. 

First of all, it is clear that Theorem \ref{th.con1} ({\bf Construction \Rmnum{1}}) is not the same as 
Theorems \ref{th.con2}, \ref{th.con3}, and \ref{th.con4} 
({\bf Constructions \Rmnum{2}}, {\bf \Rmnum{3}}, and {\bf \Rmnum{4}}). 
    We further indicate that {\bf Constructions \Rmnum{2}}, {\bf \Rmnum{3}}, and {\bf \Rmnum{4}} 
    respectively provided in Theorems \ref{th.con2}, \ref{th.con3} and \ref{th.con4} 
    are also different from each other in at least the following three regards.  
    \begin{itemize}
        \item [\rm (1)] According to Theorem \ref{th.NSC leading principalminors} 
        (resp. Algorithm \ref{alg.NSC quasi-sigma matrix}), 
        the defining matrix $LA$ used in {\bf Construction \Rmnum{2}} is usually not 
        a Toeplitz matrix, even if $A$ is chosen to be a Toeplitz matrix. 
        However, the defining matrix $A$ used in {\bf Constructions \Rmnum{3}} and {\bf \Rmnum{4}} must be a Toeplitz matrix. 
        This suggests that {\bf Construction \Rmnum{2}} is different from {\bf Constructions \Rmnum{3}} and {\bf \Rmnum{4}}.

        \item [\rm (2)] $M_{\tau}=(D_{\widehat{\tau}}\pi_e(D_{\widehat{\tau}}))\otimes M_{\tau'}$,  
        $M_{\tau}=(MQ)\otimes M_{\tau'}$ and $M_{\tau}=D\otimes M_{\tau'}$ are required in 
        {\bf Constructions \Rmnum{2}}, {\bf \Rmnum{3}}, and {\bf \Rmnum{4}}, accordingly. 
        Note that $D_{\widehat{\tau}}\pi_e(D_{\widehat{\tau}})$ and $D$ must be diagonal matrices, but $MQ$ may not be.   
        For instance, in Example \ref{exam.con3}, $MQ=\adiag(\omega^{27},\omega^{54},\omega^{27})$ is not diagonal, where $\omega$ 
        is a primitive element of $\F_{2^6}$.  
        This shows that {\bf Construction \Rmnum{3}} is different from {\bf Constructions \Rmnum{2}} and {\bf \Rmnum{4}}.

        \item [\rm (3)] If one lets $\sigma'=(\tau',\pi_e)\in \SLAut(\F_q^n)$ be the same in 
          {\bf Constructions \Rmnum{2}}, {\bf \Rmnum{3}}, and {\bf \Rmnum{4}},  
        then the input codes $\C_1, \C_2, \ldots, \C_s$ used in {\bf Constructions \Rmnum{2}} and {\bf \Rmnum{3}} are limited to 
        be $\sigma'$ self-orthogonal, but the $\sigma'$ self-orthogonality is not a necessary condition for {\bf Construction \Rmnum{4}}. 
        This shows that {\bf Construction \Rmnum{4}} is different from {\bf Constructions \Rmnum{2}} and {\bf \Rmnum{3}}. 
    \end{itemize}  
    In summary, the last three general constructions of $\sigma$ self-orthogonal matrix-product codes, 
    and hence, all these four general constructions, differ from each other. 

{
\begin{remark}\label{rem.dualgood}
    In Examples \ref{example.con2} and \ref{exam.con3}, we {obtained} some 
    $\sigma$ self-orthogonal matrix-product codes over $\F_{3^4}$ and $\F_{2^6}$ 
    and listed them in Tables \ref{tab3} and \ref{tab4}, respectively. 
    Note that their parameters cannot be compared to the code table \cite{M2022}.
    Note also that there is no general $\sigma$ self-orthogonal matrix-product code with explicit parameters constructed over finite fields as far as we know. 
    In this sense, we can say that these $\sigma$ self-orthogonal matrix-product codes have the best-known parameters.  
    On the other hand, according to \cite{C2024Arxiv}, the $\sigma$ self-orthogonal matrix-product codes constructed in this paper 
    can be employed to derive entanglement-assisted quantum codes. 
    For this case, one expects to get a $\sigma$ self-orthogonal code such that its $\sigma$ dual code has a large minimum distance.  
    In the following, we give some explicit examples.
\end{remark}

\begin{example}
    Let $q=2^3=8$ and $e=0$. Let $A\in \M(\F_{8},3)$ be an NSC Toeplitz matrix, 
    $M\in \M(\F_{8},3)$ be a monomial matrix, 
    and $D=\diag(\omega^{4}, \omega^{6},1)\in \M(\F_{8},3)$ be a diagonal matrix given by  
    $$A=\left(\begin{array}{ccc}
        1 & \omega^{6} & \omega^{3} \\
        \omega^{4} & 1 & \omega^{6} \\ 
        \omega^{6} & \omega^{4} & 1  
    \end{array}
    \right)~{\rm and}~M=\left(\begin{array}{ccc}
        \omega^{3} & 0 & 0 \\
        0 & \omega^{6} & 0 \\ 
        0 & 0 & \omega^{3}
    \end{array}
    \right),$$
    where $\omega$ is a primitive element of $\F_{8}$. 
    It can be checked that $\pi_0(A)MA=DQ$.  
    According to \cite[Theorem 3]{LCC2018} again, there exist $(\tau_ {id},\pi_0)$ ($i.e.,$ Euclidean) self-orthogonal 
    $[n,k,n+1-k]_{8}$ MDS codes for any $2\leq k\leq \lfloor \frac{n}{2} \rfloor$ and $4\leq n\leq 8$. 
    Take $\sigma=(\tau,\pi_0)$ and $\sigma'=(\tau_ {id},\pi_0)$, where $\tau$ corresponds to the monomial matrix 
    $M_{\tau}=(MQ)\otimes I_{n}=\adiag(\omega^{3},\omega^{6},\omega^{3})\otimes I_{n}$. 
    Let $\C_i=[n,k_i,n+1-k_i]_{8}$ for $2\leq k_i\leq \lfloor \frac{n}{2} \rfloor$, $4\leq n\leq 8$ and $1\leq i\leq 3$. 
    With Theorem \ref{th.con3}, we immediately get $\sigma$ self-orthogonal matrix-product codes with parameters 
    $\left[3n,k_1+k_2+k_3,\geq \min\{(4-i)(n+1-k_i)\mid 1\leq i\leq 3\}\right]_{8}$ 
    and their $\sigma$ dual codes have parameters $[3n,3n-k_1-k_2-k_3,\geq \min\{i(k_i+1)\mid 1\leq i\leq 3\}]_{8}$.
    We collect some examples of $\sigma$ self-orthogonal matrix-product codes with good $\sigma$ dual distances in Table \ref{tab5}. 
    Note that more examples can be obtained similarly. 

    \begin{table}[h]
        \centering
        \caption{{Some $\sigma$ self-orthogonal matrix-product codes with good $\sigma$ dual distances}}\label{tab5}
        \resizebox{\textwidth}{!}{
        \begin{tabular}{c|c|c|c|c}
            \hline  
            $n$ & $(k_1, k_2, k_3)$ & $\sigma$ self-orthogonal matrix-product code & $\sigma$ dual code & Best-known parameters in \cite{M2022} \\ \hline\hline
            
            $6$ & $(3,2,2)$ & $[18,7,\geq 5]_8$ & $[18,11,\geq 4]_8$ & $[18,11,6]_8$ \\ \hline 
            
            $7$ & $(2,2,2)$ & $[21,6,\geq 6]_8$ & $[21,15,\geq 3]_8$ & $[21,15,5]_8$ \\ \hline 

            $8$ & $(2,2,2)$ & $[24,6,\geq 7]_8$ & $[24,18,\geq 3]_8$ & $[24,18,5]_8$ \\ \hline 

            $8$ & $(3,2,2)$ & $[24,7,\geq 7]_8$ & $[24,17,\geq 4]_8$ & $[24,17,6]_8$ \\ \hline 

            $8$ & $(4,2,2)$ & $[24,8,\geq 7]_8$ & $[24,16,\geq 5]_8$ & $[24,16,7]_8$ \\ \hline 

        \end{tabular}}
    \end{table}
\end{example}
}

\section{Concluding remarks}\label{sec4}

In this paper, we focused on the study of $\sigma$ self-orthogonal matrix-product codes 
associated with Toeplitz matrices and presented four general constructions,  
which were based on the $\sigma'$ dual of a known $\sigma'$ dual-containing matrix-product code,  
the newly introduced NSC quasi-$\widehat{\sigma}$ matrices, and the utilization of 
certain special Toeplitz matrices, respectively. 
We also proposed a concrete construction for the NSC quasi-$\widehat{\sigma}$ matrices 
and provided an efficient algorithm to obtain such matrices by employing NSC Toeplitz matrices. 
{In particular, it has an interesting application in constructing NSC quasi-unitary matrices.
As a byproduct, we also found an attractive connection between these special Toeplitz matrices and 
$\widetilde{\tau}$-OD matrices. 
Additionally, according to \cite{C2024Arxiv}, the $\sigma$ self-orthogonal matrix-product codes constructed in this paper 
can be used to derive entanglement-assisted quantum codes.}

For future research, it would be interesting to construct more $\sigma$ self-orthogonal matrix-product codes 
for various $\sigma$ inner products over finite fields or finite rings and explore their possible applications 
in other areas.

\section*{Acknowledgments}
{The authors are very grateful to the anonymous reviewers and the Editors, Prof.
Markus Grassl and Prof. Claude Carlet, for their valuable comments that greatly improved the presentation and quality of this paper.}
The research of Yang Li and Shixin Zhu was supported by the National Natural Science Foundation of China under Grant Nos. 12171134 and U21A20428.
The research of Edgar Mart\'inez-Moro was supported by MCIN/AEI/10.13039/501100011033 and the European Union
NextGenerationEU/PRTR under Grant TED2021-130358B-I00. All the examples in this paper were computed or verified with the Magma software package.

\section*{Declarations}

\noindent\textbf{Data availability} No data are generated or analyzed during this study.  \\

\noindent\textbf{Conflict of Interest} The authors declare that there is no possible conflict of interest.

\end{sloppypar}
\end{document}